\documentclass[journal,twoside,web]{ieeecolor}
\usepackage{tmi}
\usepackage{cite}
\usepackage{amsmath,amssymb,amsfonts}
\usepackage{graphicx}
\usepackage{textcomp}
\usepackage{algorithm}
\usepackage{algpseudocode}
\usepackage{bbding}
\usepackage{makecell}
\usepackage{colortbl}
\usepackage{xcolor}
\usepackage{booktabs}
\usepackage{multirow}

\usepackage{framed}
\usepackage{color}
\usepackage{soul}
\usepackage{mdframed}
\usepackage{capt-of}
\definecolor{shadecolor}{rgb}{0.99,0.98,0.93}

\def\BibTeX{{\rm B\kern-.05em{\sc i\kern-.025em b}\kern-.08em
    T\kern-.1667em\lower.7ex\hbox{E}\kern-.125emX}}
\begin{document}
\title{Unsupervised CT Metal Artifact Reduction by Plugging Diffusion Priors in Dual Domains}
\author{Xuan Liu, Yaoqin Xie, Songhui Diao, Shan Tan, \IEEEmembership{Member, IEEE}, and Xiaokun Liang, \IEEEmembership{Member, IEEE}
\thanks{This work is partly supported by grants from the National Key Research and Develop Program of China (2023YFC2411502), National Natural Science Foundation of China (62071197, 82202954, U20A20373, U21A20480). (\textit{Corresponding author: Shan Tan and Xiaokun Liang.})}
\thanks{Xuan Liu is with the
School of Artificial Intelligence and Automation, Huazhong University of Science and Technology, Wuhan 430074, China and the Institute of Biomedical and Health Engineering, Shenzhen Institutes of Advanced Technology, Chinese Academy of Sciences, Shenzhen 518055, China (e-mail: liuxuan99@hust.edu.cn).}
\thanks{Shan Tan is with the
School of Artificial Intelligence and Automation, Huazhong University of Science and Technology, Wuhan 430074, China (e-mail: shantan@hust.edu.cn).}
\thanks{Yaoqin Xie, Songhui Diao, and Xiaokun Liang are with the Institute of Biomedical and Health Engineering, Shenzhen Institutes of Advanced Technology, Chinese Academy of Sciences, Shenzhen 518055, China (e-mail: yq.xie@siat.ac.cn, sh.diao@siat.ac.cn, xk.liang@siat.ac.cn).}
}

\maketitle

\begin{abstract}
During the process of computed tomography (CT), metallic implants often cause disruptive artifacts in the reconstructed images, impeding accurate diagnosis. Many supervised deep learning-based approaches have been proposed for metal artifact reduction (MAR). However, these methods heavily rely on training with paired simulated data, which are challenging to acquire. This limitation can lead to decreased performance when applying these methods in clinical practice. Existing unsupervised MAR methods, whether based on learning or not, typically work within a single domain, either in the image domain or the sinogram domain. In this paper, we propose an unsupervised MAR method based on the diffusion model, a generative model with a high capacity to represent data distributions. Specifically, we first train a diffusion model using CT images without metal artifacts. Subsequently, we iteratively introduce the diffusion priors in both the sinogram domain and image domain to restore the degraded portions caused by metal artifacts. Besides, we design temporally dynamic weight masks for the image-domian fusion. The dual-domain processing empowers our approach to outperform existing unsupervised MAR methods, including another MAR method based on diffusion model. The effectiveness has been qualitatively and quantitatively validated on synthetic datasets. Moreover, our method demonstrates superior visual results among both supervised and unsupervised methods on clinical datasets. Codes are available in \underline
{github.com/DeepXuan/DuDoDp-MAR}.   
\end{abstract}

\begin{IEEEkeywords}
Computed tomography, Metal artifact reduction, Diffusion model, Unsupervised learning
\end{IEEEkeywords}

\section{Introduction}
\label{sec:introduction}
\IEEEPARstart{C}{omputed} tomography serves as an indispensable tool for both diagnosis and treatment planning. However, the presence of metallic implants in patients can introduce severe artifacts into the reconstructed CT images~\cite{de1998metal, park2018ct}, thereby compromising the integrity of subsequent diagnostic evaluations. Specifically, metallic implants distort portions of the sinogram, leading to anomalous values. Consequently, CT image reconstructed from such perturbed sinogram always exhibit disruptive artifacts in non-metal regions~\cite{kalender1987reduction}. To mitigate these metal-affected artifacts, plenty of algorithms have been developed for use in instances where the sinogram is contaminated by metallic implants.

The most commonly employed traditional methods for metal artifact reduction (MAR) are based on sinogram inpainting, which remove and inpaint the metal-affected regions in the sinogram before performing the reconstruction process. Kalender et al. proposed to fill the hollow sinogram using linear interpolation (LI)~\cite{kalender1987reduction}. Meyer proposed normalized metal artifact reduction (NMAR) to replace the metal-affected regions in sinogram with the forward projection of a prior image~\cite{meyer2010normalized}. Although LI and NMAR are both simple and effective MAR methods, the reconstructed images may still contain artifacts due to the potential discontinuity at the edges of the completed sinogram. 

With the rapid development of deep learning techniques over the past decade, numerous MAR methods based on deep learning have been proposed, exhibiting significant improvements compared to traditional methods. The majority of deep learning-based MAR methods are supervised, relying on paired metal artifact CT and clean CT data to train neural networks~\cite{lin2019dudonet,wang2021indudonet,wang2022ada,zhou2022dudodr,wang2023indudonet+}. Deep learning-based MAR methods can work in sinogram domain, image domain, or both. Park et al. proposed to use a convolutional neural network (CNN) to  repair inconsistent sinogram caused by beam hardening~\cite{park2018ct}. Zhang et al. proposed CNNMAR that produced a prior image by a CNN then inpainted the metal-affected sinogram with the forward projection of the prior image~\cite{zhang2018convolutional}. Besides, post-processing deep neural networks attempt to directly learn the mapping from metal artifact CT images to clean CT images. Huang et al. conducted MAR for cervical CT images with deep residual learning~\cite{huang2018metal}. Wang et al. used generative adversarial networks (GAN)~\cite{goodfellow2020generative} for MAR of ear CT images~\cite{wang2018conditional}. By utilizing information from both the sinogram domain and the image domain, some dual-domain networks have been proposed to simultaneously restore sinogram consistency and enhance CT images. Lin et al. proposed an end-to-end trainable dual domain network (DuDoNet)~\cite{lin2019dudonet}. Zhou et al. proposed a dual-domain data consistent recurrent network (DuDoRNet)~\cite{zhou2022dudodr}. Wang et al. proposed an interpretable dual domain network (InDuDoNet)~\cite{wang2021indudonet} and its improved version, InDuDoNet+\cite{wang2023indudonet+}. The dual-domain networks process both of the sinogram data and image data successively or recurrently, and achieve better performance compared with single-domain networks. 

However, training an MAR neural network in a supervised manner requires large amounts of paired CT data with and without metal artifacts, which are practically unattainable in clinical. As a result, supervised methods often have to be trained exclusively on simulated data, which can lead to suboptimal results in clinical applications due to domain gaps. Therefore, some unsupervised or weakly supervised methods have been proposed to address the issue of lacking paired training data. Liao et al. introduced the first unsupervised learning approach to MAR named artifact disentanglement network (ADN), which realized translation between metal-affected CT images and clean CT images~\cite{liao2019adn}. Lyu et al. proposed U-DuDoNet, a dual-domain network trained on unpaired metal artifact CT and clean CT images for MAR~\cite{lyu2021u}. Du et al. proposed a novel MAR method based on unsupervised domain adaptation (UDA) called UDAMAR to tackle the domain gap problem~\cite{du2023deep}. In addition, certain general unsupervised image-to-image translation methods can also be employed for MAR, such as CycleGAN~\cite{zhu2017unpaired}, MUNIT~\cite{huang2018multimodal}. These prevalent unsupervised MAR methods work solely within the image domain. However, due to the severe and unstructured nature of metal artifacts in the image domain, the effectiveness of these methods, particularly in terms of visual quality, is often limited. Wu et al. proposed an unsupervised MAR method based on implicit neural representation, which is highly effective but extremely time-consuming~\cite{wu2023unsupervised}. Song et al. proposed a novel unsupervised approach for MAR, wherein they formulated the task as a linear inverse problem and utilized a pre-trained score model~\cite{song2021scorebased} as a regularizer~\cite{song2022solving}. Specifically, they iteratively performed inpainting on the hollow regions in the sinogram domain using prior images from the score model. The method is insightful but we consider it doesn't fully leverage the powerful capabilities of score models because the combination of the priors provided by the score model and the known portions only occurs in the sinogram domain.

In order to address the limitations of existing unsupervised MAR methods, we proposed a novel method for MAR that utilize the priors embedded within a pre-trained diffusion model~\cite{ho2020denoising} in both sinogram domain and image domain.

In addressing the challenge of MAR in CT imaging, the modeling of image priors is of paramount importance. The diffusion model, renowned for its prowess as a generative model, excels in producing high-quality CT images free from artifacts. Moreover, it iteratively provides analytical probability density functions of the generated data. This characteristic is particularly beneficial for our study as it aids in the accurate and efficient reconstruction of images. Compared to other commonly used generative models, the diffusion model can generate more realistic images than variation autoencoders (VAEs)~\cite{kingma2013auto} and flows-based models~\cite{dinh2014nice,dinh2016density,2018Glow}, while also possessing explicit distribution representation capabilities that generative adversarial networks (GANs)~\cite{goodfellow2020generative} lack. Therefore, the diffusion model is well-suited to provide priors for solving the MAR problem. 

Specifically, we first train a diffusion model with large amounts of CT images without metal artifacts for preparation. The well-trained diffusion model is not only capable of generating visually realistic clean CT images but can also represent its distribution using analytical probability density functions. Subsequently, we extract the metal-affected portions from the sinogram and model MAR as a reconstruction problem with partial measurement missing. This problem is solved iteratively. At each timestep of the diffusion generation process, we first fuse the output of diffusion model and the known sinogram in the sinogram domain. Due to the discontinuity between the prior sinogram provided by the diffusion model and the known sinogram, the reconstructed image might introduce new artifacts in the image domain. Therefore, we reintroduce the diffusion priors in the image domain to mitigate this issue. Besides, to ensure a balance between likelihood and prior, we further design weight masks for the image-domain fusion and enhance the performance.     

We name this \textbf{du}al \textbf{do}main method with \textbf{d}iffusion \textbf{p}riors as \textbf{DuDoDp}. 
Our contributions can be summarized as follows:
\begin{enumerate}[]
	\item We present an MAR method that combines diffusion priors and known measurements within a dual-domain framework to fully leverage the information from both the sinogram domain and the image domain.
	\item We propose to fuse three kinds of images at each timestep, which are prior images from diffusion model, reconstructed images from inpainted sinogram, and original metal-artifact images. The combination of these images corrects the artifacts caused by sole sinogram inpainting.      
	\item We design dynamic weight masks to control the addition of diffusion priors. According to the iterative generation nature of diffusion, we dynamically adjust the masks over timesteps to achieve enhanced results.
	\item The proposed method is validated in both synthetic and clinical datasets, demonstrating its potential for clinical applications. 
\end{enumerate}

\section{Method}
\subsection{Preliminary}
\begin{figure*} \renewcommand{\thefigure}{1}
	\centering\includegraphics[width=0.95\textwidth]{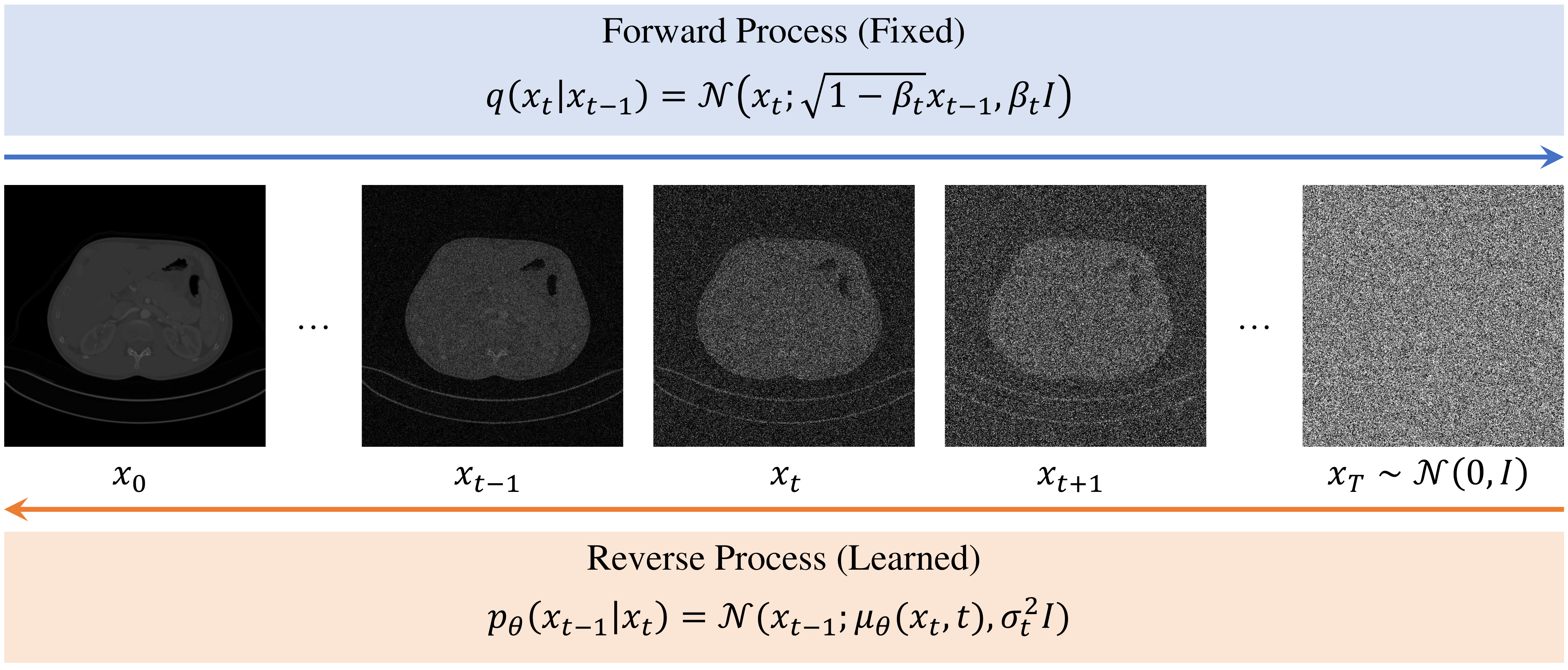}
	\caption{A typical diffusion model consists of a forward process and a reverse process. The forward process gradually adds noise to clean images, and the reverse process generates clean images from Gaussian noise.} \label{fig1_}
\end{figure*} 

\subsubsection{Diffusion Model} The diffusion model (short for denoising diffusion probabilistic model) is a kind of deep generative model~\cite{ho2020denoising,song2021scorebased,croitoru2023diffusion}. In general, a diffusion model consists of a forward process and a reverse process. The forward process iteratively transforms clean data into Gaussian noise, and the reverse process employs neural networks to learn the inverse of the forward process.

Given clean data $x_0$, the forward process iteratively samples $x_1, x_2, \ldots, x_T$ through a conditional distribution $q\left(x_t \mid x_{t-1}\right)$, as shown in the upper parts of Fig. \ref{fig1_}. where $T$ denotes the total number of timesteps. $q\left(x_t \mid x_{t-1}\right)$ is defined as,
\begin{equation}
	q\left(x_t \mid x_{t-1}\right) :=\mathcal{N}\left(x_t ; \sqrt{1-\beta_t} x_{t-1}, \beta_t I\right),
	\quad t=1, \ldots, T, 
\end{equation}
where $\mathcal{N}\left(\cdot; \mu, \Sigma \right)$ denotes a Gaussian distribution with mean $\mu$ and covariance matrix $\Sigma$, and $\{\beta_1, \ldots, \beta_T\}$ is a fixed variance schedule. In our experiments, we set $\beta_1=1.0\times10^{-4}$, $\beta_T=2.0\times10^{-2}$, and the intermediate values increase linearly. The forward process defined in this way results in increasing noise on $x_t$ as t grows. When $t$ equals $T$, $x_T$ is almost entirely Gaussian noise, that's to say, $x_T\sim \mathcal{N}\left(0, I\right)$. According to the definition of the forward process, we can derive the following two conditional distributions, which will be used in subsequent methods,  
\begin{equation}
	q\left(x_t \mid x_0\right)=\mathcal{N}\left(x_t ; \sqrt{\bar{\alpha}_t} x_0,\left(1-\bar{\alpha}_t\right) I\right),  \label{eq2__}
\end{equation}

\begin{equation}
	q\left(x_{t-1} \mid x_t, x_0\right)=\mathcal{N}\left(x_{t-1} ; \tilde{\mu}_t\left(x_t, x_0\right), \tilde{\beta}_t I\right), \label{eq3__}
\end{equation}

where,
\begin{align}
	&\bar{\alpha}_t:=\prod_{i=1}^t (1-\beta_i),\\ 
	&\tilde{\mu}_t\left(x_t, x_0\right):=\frac{\sqrt{\bar{\alpha}_{t-1}} \beta_t}{1-\bar{\alpha}_t} x_0+\frac{\sqrt{\alpha_t}\left(1-\bar{\alpha}_{t-1}\right)}{1-\bar{\alpha}_t} x_t, \label{eq5__}\\
	&\tilde{\beta}_t:=\frac{(1-\bar{\alpha}_{t-1})}{1-\bar{\alpha}_t} \beta_t. \label{eq6__}
\end{align}

The reverse process aims to generate clean image $x_0$ from Gaussian noise $x_T$, which can be understood as the inverse of the forward process, as shown in the lower parts of Fig. \ref{fig1_}. The reverse process samples $x_{t-1}$ given $x_t$ through a conditional distribution $p_\theta\left(x_{t-1} \mid x_t\right)$ with learnable parameters $\theta$, where
\begin{equation}
	p_\theta\left(x_{t-1} \mid x_t\right):=\mathcal{N}\left(x_{t-1} ; \mu_\theta\left(x_t, t\right), \sigma^2_{t}I\right),  \quad t=T, \ldots, 1, \label{eq7__}
\end{equation}  
where $\mu_\theta\left(x_t, t\right)$ is predicted by trainable neural networks, and $\sigma^2_{t}$ is always set as constants, and the iteration starts from $x_T\sim \mathcal{N}\left(0, I\right)$. 

To train the reverse process such that the generated data matches the known data distribution, the ideal objective would be the maximum likelihood of training data, $p_{\theta}(x_0)$. However,  $p_{\theta}(x_0)$ is intractable because it is obtained in an iterative way. Therefore, the variational bound of the negative log likelihood is always used as diffusion models' training loss~\cite{ho2020denoising,croitoru2023diffusion}, which is, 
\begin{equation}
	\begin{split}
	L_{vb}(\theta, x_0)=&D_{\mathrm{KL}}\left(q\left(x_T \mid x_0\right) \| p\left(x_T\right)\right) \\
	+&\sum_{t>1}D_{\mathrm{KL}}\left(q\left(x_{t-1} \mid x_t, x_0\right) \| p_\theta\left(x_{t-1} \mid x_t\right)\right)\\
	-&p_\theta\left(x_{0} \mid x_1\right) \geq -\log p_\theta\left(x_0\right),
	\end{split} 
\end{equation}  
where the latter two terms are related to $\theta$, $D_{\mathrm{KL}}$ denotes the KL divergence of two distribution, $q\left(x_{t-1} \mid x_t, x_0\right)$ and $p_\theta\left(x_{t-1} \mid x_t\right)$ can be obtained from (\ref{eq3__}) and (\ref{eq7__}). The training only utilized a single set of clean data $\{x_0^{i}\}_{i=1,\ldots,N}$. 

\subsubsection{$\epsilon$-Prediction and $x_0$-Prediction Diffusion Model} Generally, a diffusion model doesn't directly predict the mean values $\mu_\theta\left(x_t, t\right)$ in (\ref{eq7__}) at each timestep. Alternatively, it's more common to predict the added noise $\epsilon_t$ or the initial image $x_0$ from $x_t$, where $x_t=\sqrt{\bar{\alpha}_t} x_0+\sqrt{1-\bar{\alpha}_t}\epsilon_t$. The relationship between the predicted noise $\epsilon_{\theta}(x_t, t)$ and $\mu_\theta\left(x_t, t\right)$ is~\cite{ho2020denoising},
\begin{equation}
	\begin{split}
	\mu_\theta\left(\mathbf{x}_t, t\right)&=\tilde{\mu}_t\left(\mathbf{x}_t, \frac{1}{\sqrt{\bar{\alpha}_t}}\left(x_t-\sqrt{1-\bar{\alpha}_t} \epsilon_\theta\left(x_t\right)\right)\right)\\
	&=\frac{1}{\sqrt{\alpha_t}}\left(x_t-\frac{\beta_t}{\sqrt{1-\bar{\alpha}_t}} \epsilon_\theta\left(x_t, t\right)\right),
	\end{split}
\end{equation}    
where $\tilde{\mu}_t$ is from (\ref{eq5__}). Predicting $\epsilon_{\theta}(x_t, t)$ from $x_t$ makes the training of diffusion models more concise to implement~\cite{ho2020denoising}.  

Naturally, we can also predict the latent clean images $x_0$ from $x_t$. Let the prediction of $x_0$ be $f_\theta\left(x_t, t\right)$, we have, 
\begin{equation}
	f_\theta\left(x_t, t\right) = (x_t - \sqrt{1-\bar{\alpha}_t}\epsilon_{\theta}(x_t, t)) / \sqrt{\bar{\alpha}_t}. \label{eq10__}
\end{equation}
Under this setting, the inference process of the diffusion model in (\ref{eq7__}) can be rewritten as follows~\cite{song2020denoising},
\begin{equation}
	p_\theta\left(x_{t-1} \mid x_t\right) = q\left(x_{t-1} \mid x_t, f_\theta\left(x_t, t\right)\right), \label{eq11__}
\end{equation}
where $q\left(x_{t-1} \mid x_t, f_\theta\left(x_t, t\right)\right)$ is calculated according to (\ref{eq3__}).

Regardless of whether predicting $\mu_\theta$, $\epsilon_\theta$, or $f_\theta$, the principles of a diffusion model and the meaning of its training loss remain unchanged. The only difference lies in the interpretation of the neural network outputs. 
\\
\begin{figure}
	\centering\includegraphics[width=0.45\textwidth]{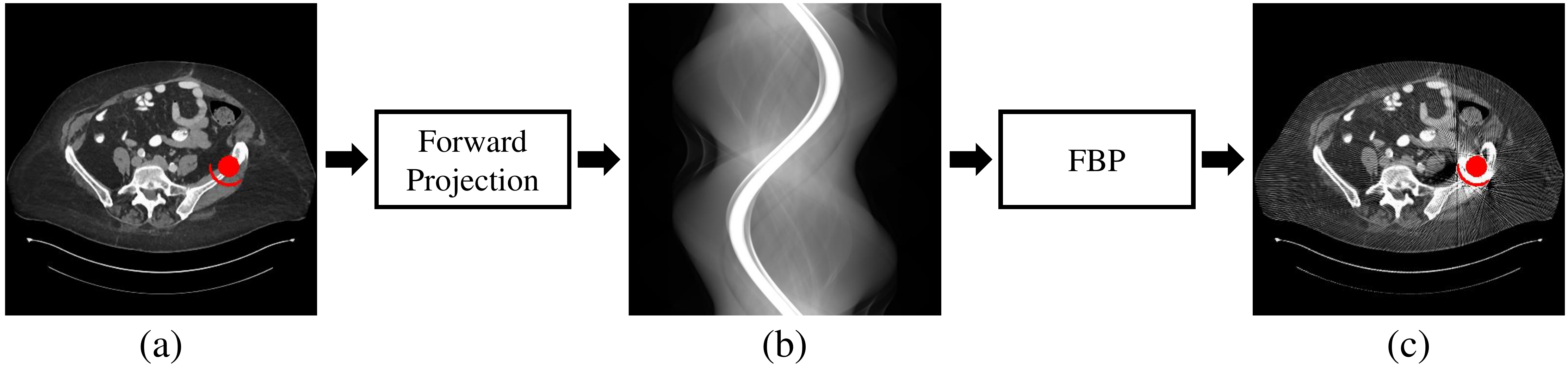}
	\caption{The presence of metallic implants (a) can lead to various physical effects on the sinogram (b), resulting in significant metal artifacts in the reconstructed images (c).} \label{fig2_}
\end{figure} 

\subsubsection{Mathematical Model of MAR problem} When a patient has metallic implants within their body (as shown in Fig. \ref{fig2_}(a)), during a CT scan, parts of the sinogram will be affected by various physical phenomena, such as beam hardening, photon starvation~\cite{zhang2018convolutional} (as shown in Fig. \ref{fig2_}(b)). If reconstruction is carried out without proper processing, it can result in significant metal artifacts in the non-metal regions of the reconstructed image (as shown in Fig. \ref{fig2_}(c)). 

Metal artifact reduction (MAR) aims to remove the metal artifacts in non-metal regions of the reconstructed CT image. One direct way of MAR is to correct the physical effects in the metal-affected sinogram~\cite{hsieh2000iterative,kachelriess2001generalized,zhang2010beam,park2015metal}. However, these methods can't achieve satisfactory performance because metallic implants can lead to strong errors in the sinogram. Another approach is to consider the metal-affected portions in the sinogram as missing and then design a method to inpaint the missing parts~\cite{kalender1987reduction,meyer2010normalized,zhang2018convolutional,song2022solving,wang2023indudonet+}. We choose this method to model the MAR problem. 

Let $s_0$ be a CT sinogram affected by metal, where the affected region is represented by a boolean mask $\mathcal{M}_s$. The elements of $\mathcal{M}_s$ in the affected parts are set to 1, while elements in the unaffected portion are set to 0. Typically, $\mathcal{M}_s$ is known because the metallic part of the reconstructed image can be easily segmented using thresholding and then forward-projected to identify the affected regions in the sinogram~\cite{wang2021indudonet,wang2023indudonet+}. In our algorithm, $\mathcal{M}_s$ is treated as a known matrix. Assuming $x_0$ is the latent artifact-free reconstructed image to solve, there is,   
\begin{equation}
	(1-\mathcal{M}_s)\odot \mathrm{FP}(x_0)=(1-\mathcal{M}_s)\odot s_0, \label{eq12__}
\end{equation}
where $\mathrm{FP}(\cdot)$ denotes the forward projection function of the CT system, $\odot$ denotes the Hadamard product.

\subsection{MAR with Dual Domain Diffusion Priors} \label{sec2b}
\begin{figure*}[t]
	\centering\includegraphics[width=0.95\textwidth]{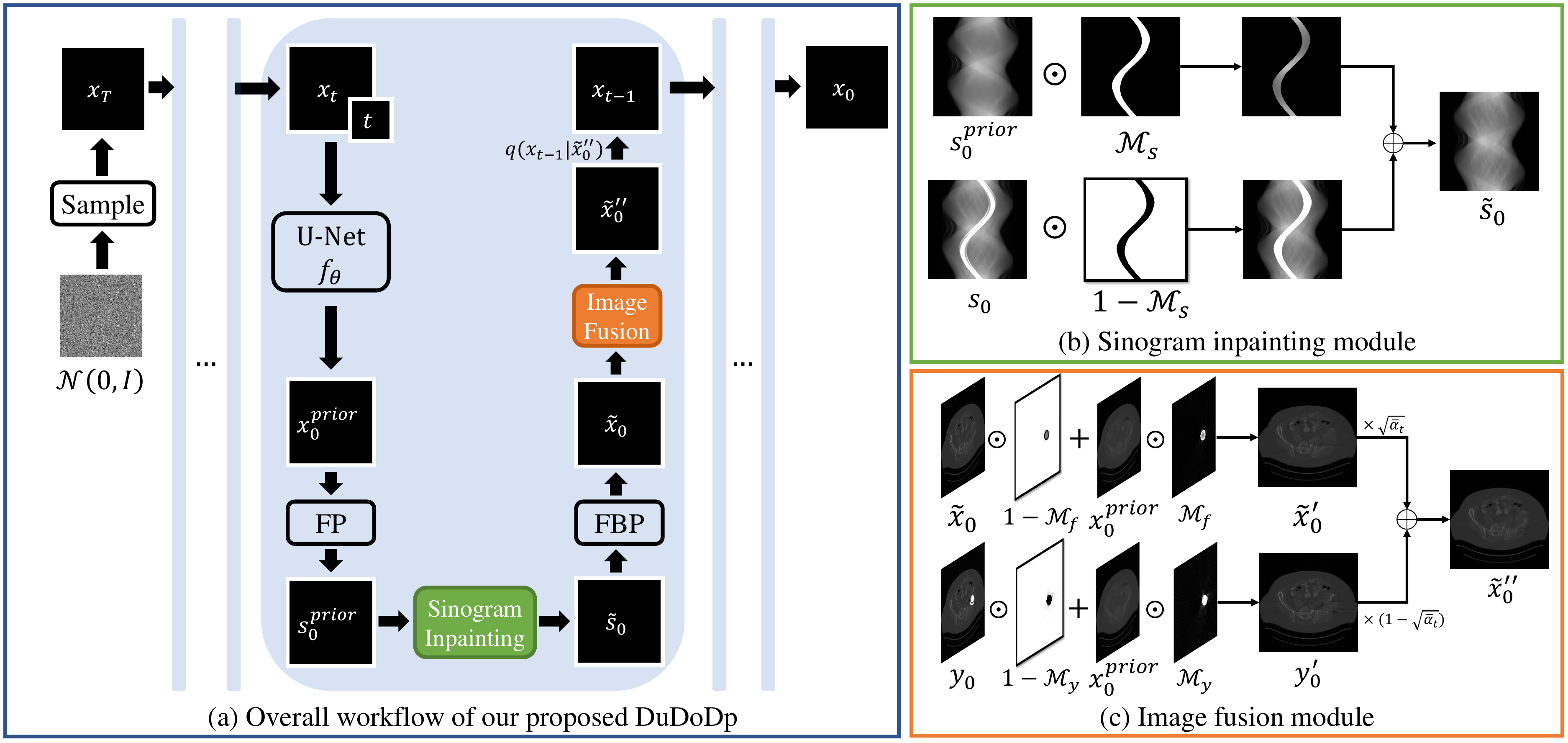}
	\caption{(a) illustrates the overall workflow of our MAR method, DuDoDp, where each blue box shows one iteration to calculate $x_{t-1}$ from $x_t$ while the initial $x_T$ is sampled form a unit Gaussian distribution. (b) and (c) respectively illustrate the sinogram inpainting module and image fusion module in (a).} \label{fig3_}
\end{figure*}

In this section, we introduce DuDoDp, an MAR method with dual-domain diffusion priors. Firstly, in Section \ref{secb1}, we introduce the diffusion priors into the MAR problem by iteratively solving multiple optimization problems. This incorporation is equivalent to employing the prior images provided by the diffusion model for inpainting in the sinogram domain. To alleviate the artifacts arising from the discontinuity in the inpainted sinogram, we propose further integration of the diffusion priors in the image domain, as outlined in Section \ref{secb2}. In Section \ref{secb3}, we design weight masks for image domain fusion, and considering the iterative nature of diffusion, dynamic masks are employed at different timesteps. The overall algorithm workflow is illustrated in Fig. \ref{fig3_}.  

\subsubsection{Introduction of Diffusion Priors for MAR} \label{secb1}
	To solve the MAR problem, one way is to model the prior distribution of $x_0$, $\mathcal{R}(x_0)$, then calculate a solution that meets both the likelihood constraint present in (\ref{eq12__}) and follow the prior distribution $\mathcal{R}(x_0)$.
	
	As a powerful generative model, the diffusion model can generate highly realistic and high-quality images, demonstrating its capability to represent data distributions. Therefore, we first train a diffusion model $p_\theta$ that can generate artifact-free CT images. The training scheme is based on Section II-A.1, where only a set of clean CT images $\{x_0^{i}\}_{i=1,\ldots,N}$ are used for training. 
	
	The diffusion model does not provide an analytical form for the distribution of $x_0$ directly; instead, it iteratively obtains the data distribution at each timestep $t-1$ from timestep $t$ according to (\ref{eq11__}). Therefore, we can also solve the MAR problem iteratively. Initially, we sample $x_T$ from a normal distribution $\mathcal{N}\left(0, I\right)$, and then, given $x_t$, obtain the prior and likelihood terms for $x_{t-1}$.
	
	Firstly, given $x_t$, \textbf{the prior distribution} of $x_{t-1}$ is,
	\begin{equation}
		x_{t-1}\sim p_\theta\left(x_{t-1} \mid x_t\right) = \mathcal{N}\left(x_{t-1} ; \tilde{\mu}_t\left(x_t, f_\theta\left(x_t, t\right)\right), \tilde{\beta}_t I\right), \label{eq13__}
	\end{equation}
	where $f_\theta\left(x_t, t\right)$ is the predicted initial image of diffusion model at timestep $t$ (see \ref{eq10__}), $\tilde{\mu}_t$ and $\tilde{\beta}_t$ are defined in (\ref{eq5__}) and (\ref{eq6__}).
	
	Secondly, the likelihood of $x_{t-1}$ can be derived in an indirect way. Assume $x_{t-1}$ can be derived from a latent clean image $\tilde{x}_0^{t-1}$ in the following way,
	\begin{equation}
		x_{t-1} = \tilde{\mu}_t\left(x_t, \tilde{x}_0^{t-1}\right)+\sqrt{\tilde{\beta}_{t}} \epsilon, \label{eq14__}
	\end{equation}
	where $\epsilon\sim\mathcal{N}(0, I)$. This is in accordance with (\ref{eq3__}). \textbf{The likelihood} of $x_{t-1}$ is then modeled by $\tilde{x}_0^{t-1}$ following (\ref{eq12__}),
	\begin{equation}
		(1-\mathcal{M}_s)\odot \mathrm{FP}(\tilde{x}_0^{t-1})=(1-\mathcal{M}_s)\odot s_0. \label{eq15__}
	\end{equation}
	At the same time, Eq. (\ref{eq13__}) can be expressed in another form by combining (\ref{eq14__}),
	\begin{equation}
	\begin{split}
		\left[\tilde{\mu}_t\left(x_t, \tilde{x}_0^{t-1}\right)+\sqrt{\tilde{\beta}_{t}} \epsilon\right]\sim & p_\theta\left(x_{t-1} \mid x_t\right) \\
		= &\mathcal{N}\left(x_{t-1} ; \tilde{\mu}_t\left(x_t, f_\theta\left(x_t, t\right)\right), \tilde{\beta}_t I\right). 
	\end{split} \label{eq16__}
	\end{equation}
	
	Now, to obtain a solution of $x_{t-1}$, we can first calculate a $\tilde{x}_0^{t-1}$ that follows the distribution in (\ref{eq16__}) and meet the likelihood in (\ref{eq15__}). Eq. (\ref{eq15__}) illustrates that the forward projection of $\tilde{x}_0^{t-1}$ should be the same with $s_0$ in the unaffected regions, while (\ref{eq16__}) illustrates that $\tilde{x}_0^{t-1}$ should be as similar as $f_\theta\left(x_t, t\right)$. Based on this, we obtain a solution for $\tilde{x}_0^{t-1}$ that might not be optimal but easy to compute, which is,
	\begin{equation}
		\tilde{x}_0^{t-1}=\mathrm{FBP}(\tilde{s}_0^{t-1}), \label{eq17__}
	\end{equation}
	where,
	\begin{equation}
		\tilde{s}_0^{t-1}= (1-\mathcal{M}_s)\odot s_0 + \mathcal{M}_s \odot \mathrm{FP}(f_\theta(x_t, t)). \label{eq18__}
	\end{equation}
	$\mathrm{FBP}(\cdot)$ denotes filtered back-projection algorithm for CT reconstruction, which is approximately the inverse function of $\mathrm{FP}(\cdot)$.
	
	Through the above-mentioned approach, at each timestep, known $x_t$ and $s_0$, we can calculate $\tilde{x}_0^{t-1}$ according to (\ref{eq17__}-\ref{eq18__}) then calculate $x_{t-1}$ with (\ref{eq14__})\footnote{In practical experiments, we achieve better results by using conditional probability sampling as outlined in (\ref{eq2__}) to obtain ${x}_{t-1}$, which doesn't involve $x_t$ again.}. At this point, we have integrated the diffusion priors into the MAR framework. Eq. (\ref{eq18__}) corresponds to the sinogram inpainting module in our method.

\subsubsection{Fusion Strategy in Image Domain} \label{secb2}
\begin{figure}[htbp]
	\centering\includegraphics[width=0.48\textwidth]{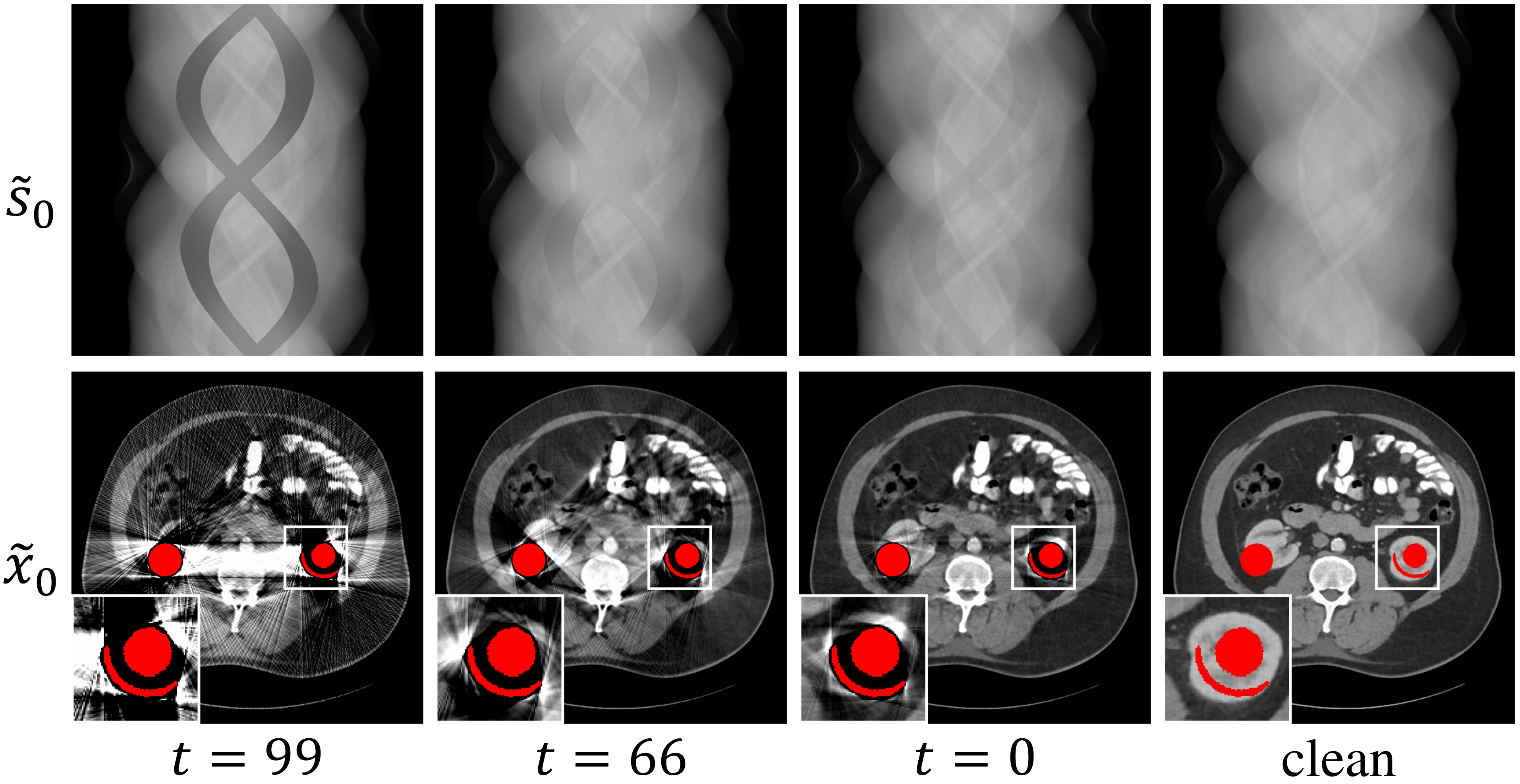}
	\caption{The intermediate results (T=100) using the diffusion priors only in the sinogram domain is shown. The red parts indicate metallic implants. It can be observed that, in the early iterations, the discontinuity of inpainted sinogram lead to artifacts beyond the metal regions, and part of these artifacts persists even after the completion of iterations. The display window of CT images is [-175, 275] HU.} \label{fig2}
\end{figure}

Although solving $x_{t-1}$ using the methods outlined in (\ref{eq17__}) and (\ref{eq18__}) is efficient, the prior sinogram $\mathcal{M}_s\odot\mathrm{FP}(f_\theta(x_t, t))$ provided by the diffusion model exhibits discontinuity with the known sinogram $(1-\mathcal{M}_s)\odot s_0$. This leads to the introduction of new artifacts when using FBP reconstruction, especially around the metal regions, as depicted in Fig. \ref{fig2}. 

Therefore, we propose refining the reconstructed image $\tilde{x}_0^{t-1}$ using the diffusion priors in the image domain. Firstly, we add the diffusion-predicted $x_0$, as well as $f_\theta(x_t, t)$, to $\tilde{x}_0^{t-1}$  at timestep $t-1$,
\begin{equation}
	\tilde{x}_0^{'t-1} = \mathcal{M}_f\odot f_\theta(x_t, t) + (1-\mathcal{M}_f)\odot \tilde{x}_0^{t-1}, \label{eq22_}
\end{equation}
where $\mathcal{M}_f$ denotes the weight mask for diffusion prior $f_\theta(x_t, t)$, which help correct the unnatural portions of $\tilde{x}_0^{t-1}$. 

In addition, we propose further fusing the original metal-artifact image $y_0$ into $\tilde{x}_0^{'t-1}$, where $y_0=\mathrm{FBP}(s_0)$. This can provide additional likelihood information. The specific approach involves,
\begin{align}
	&y_0^{'}=\mathcal{M}_y\odot f_\theta(x_t, t) + (1-\mathcal{M}_y)\odot {y}_0, \label{eq24_}\\ 
	&{\tilde{x}}_0^{''t-1} = \sqrt{\bar{\alpha}_t}\tilde{x}_0^{'t-1} + (1-\sqrt{\bar{\alpha}_t}) y_0^{'}. \label{eq25_}
\end{align}
Eq. (\ref{eq24_}) demonstrates the fusion of the metal-artifact image and the diffusion prior image using a certain weight mask $\mathcal{M}_y$. Eq. (\ref{eq25_}) illustrates the combination of the metal-artifact image that has been merged with the prior (as in (\ref{eq24_})) and the prior-enhanced reconstructed image (from (\ref{eq22_})) using a specific weight. 

In (\ref{eq25_}), $\tilde{x}_0^{'t-1}$ and $y_0^{'}$ represent the fusion of the inpainted sinogram-reconstructed image with the prior image and the fusion of the metal artifact image with the prior image, respectively. To merge these two images, we empirically choose $\sqrt{\bar{\alpha}_t}$ and $(1-\sqrt{\bar{\alpha}_t})$ as their weights. $\bar{\alpha}_t$ is a hyper-parameter of the diffusion model; as $t$ decreases, $\bar{\alpha}_t$ approaches 1, and as $t$ increases, $\bar{\alpha}_t$ approaches 0. The rationale behind this is that, as the iterations progress, the contribution of the metal artifact image decreases, since as the target image becomes cleaner, the metal artifact image can provide less helpful information. The experimental results in Table \ref{tab2} demonstrate the effectiveness of this fusion method.

The algorithm for the image-domain fusion described above is depicted in Fig. \ref{fig3_}(c). In Section IV.A, We verify the performance enhancement brought by introducing image-domain fusion strategies to our method.
\subsubsection{Temporally Dynamic Weight Masks} \label{secb3}
\begin{figure}[htbp]
	\centering\includegraphics[width=0.48\textwidth]{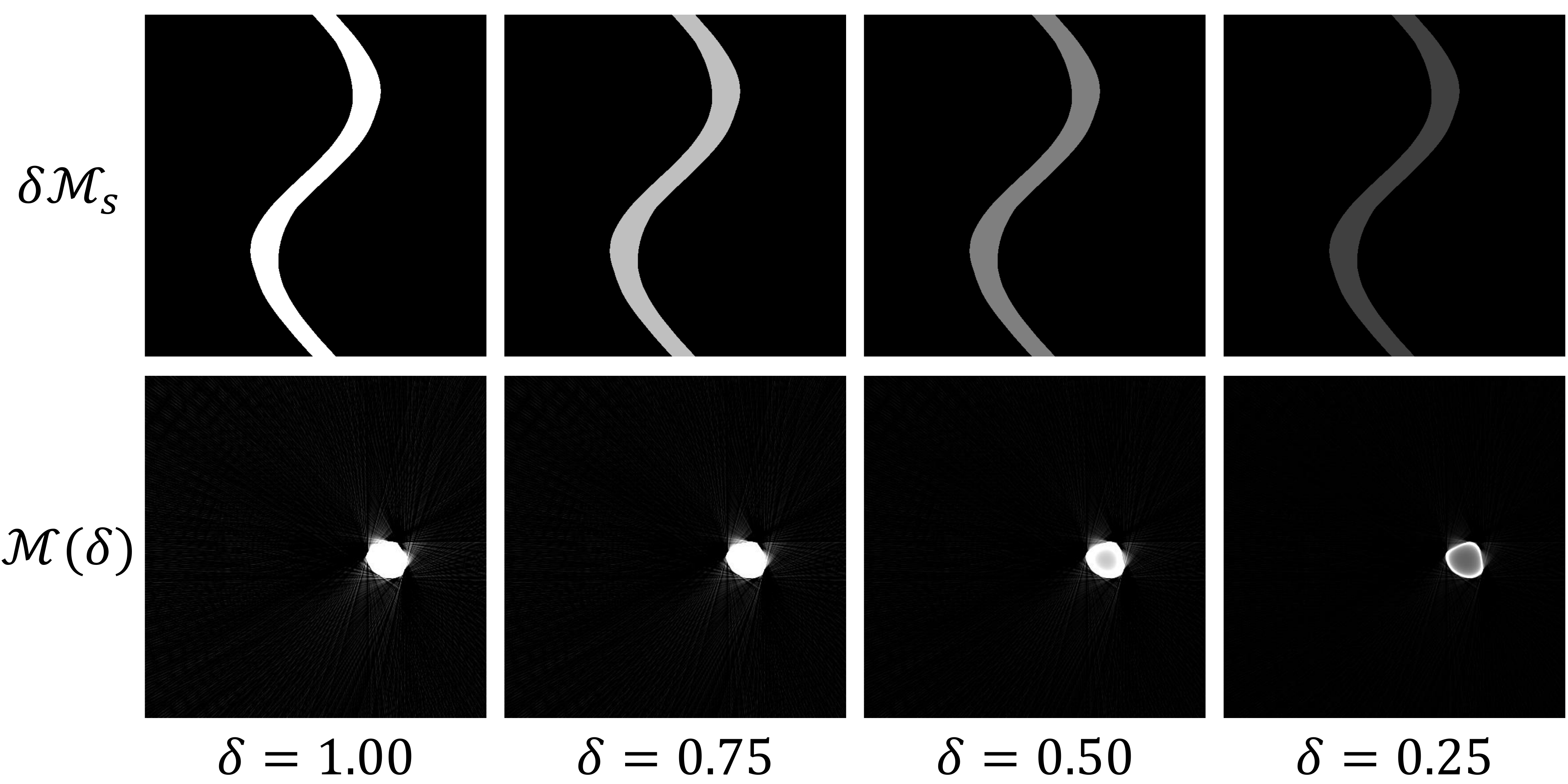}
	\caption{Binary sinogram corresponding to different $\delta$ values and their reconstructed weight masks. With smaller $\delta$ values, the resulting weight masks are generally smaller (darker in the images).} \label{fig3}
\end{figure}

Selecting proper $\mathcal{M}_f$ and $\mathcal{M}_y$ is challenging. Empirically, we propose to construct the weight mask utilizing the reconstructed results from binary sinogram as,
\begin{equation}
	\mathcal{M}(\delta) = \mathrm{FBP}( \delta\mathcal{M}_s),
\end{equation}
where $\delta$ means the value filled in the metal-affected regions while other regions are filled with zeros. Generally, as $\delta$ increases, the overall values of the reconstructed weight mask also increase, as illustrated in Fig. \ref{fig3}\footnote{Before reconstruction, the values of the sinogram are multiplied by 4.0, which corresponds to the maximum sinogram value in our experiments.}. 

For $\mathcal{M}_y$, using a fixed $\delta$ is reasonable since in (\ref{eq24_}) $y_0$ remains unchanged. Therefore, we set it as,
\begin{equation}
	\mathcal{M}_y = \mathcal{M}(\delta_y).
\end{equation} 

However, for $\mathcal{M}_f$, at different timesteps, differences between the known sinogram $(1-\mathcal{M}_s)\odot s_0$, and prior sinogram $\mathcal{M}_s\odot\mathrm{FP}(f_\theta(x_t, t))$ will cause the variations in image domain artifacts. Generally, as the value of timestep $t$ decreases, the generation of the diffusion model becomes more realistic, resulting in reduced discontinuity in inpainted sinogram and subsequently minimizing artifacts in the image domain. Therefore, designing a weight mask that varies with the timestep $t$ is more reasonable,
\begin{equation}
	\mathcal{M}_f(t) = \mathcal{M}(\delta(t)),
\end{equation}  
where $\delta(t)$ is a value varying with $t$. Then, replace $\mathcal{M}_f$ in (\ref{eq22_}) with $\mathcal{M}_f(t)$ to achieve a more accurate fusion result. In Section \ref{sec4a2}, we validate the positive impact of the temporally dynamic $\mathcal{M}_f$ on the MAR results.
\section{Implementation Details}
\subsection{Datasets}
\subsubsection{Training Dataset}
To train our diffusion model, we use the CT images without metal-artifacts from the DeepLesion dataset~\cite{yan2018deep}. Excluding the 200 CT images reserved for testing, there are a total of 927,802 512$\times$512 CT images used for training. 

\subsubsection{Synthetic Test Dataset}
We firstly evaluate our proposed methods as well as the comparison methods on a synthetic dataset, where 200 clean CT images from DeepLesion dataset are chosen to simulate metal-artifact CT images. 10 metal masks are individually implanted into each CT image, followed by using fan-beam projection simulation to obtain the metal-affected sinogram. In total, 2000 sets of CT sinogram and images affected by metal artifacts are obtained for evaluating the MAR methods. The masks are from CNNMAR~\cite{zhang2018convolutional} and the sizes of the 10 metal masks are [2061, 890, 881, 451, 254, 124, 118, 112, 53, 35] in pixels. The fan-beam geometry follows a previous work of Yu et al.~\cite{yu2020deep}. During the fan-beam projection process, the CT images were first resized to $416\times416$. Then, a total of 640 projections were evenly taken within the angle range of 0 to $2\pi$. The detector has 641 bins, resulting in a projection size of $640\times641$. During the simulation of metal artifacts, several effects are taken into consideration, including polychromatic X-ray, partial volume effect, beam hardening, and Poisson noise. This synthetic dataset is widely used in previous works~\cite{lin2019dudonet,liao2019adn,zhou2022dudodr,wang2021indudonet,wang2022ada,wang2023indudonet+}.
\subsubsection{Clinical Test Datset}

We further conduct test of our method on the CTPelvic1K dataset~\cite{liu2021deep}. The sub-dateset clinical-metal of CTPelvic1K contains postoperative CT images with metal artifacts. The clinical CT images are firstly resized to $416\times416$. Then, following the same projection principle with synthetic DeepLesion data, we acquire the sinogram of the clinical data. We segment the metallic implants using a threshold of 2500 HU and obtain the metal trace in the sinogram through forward projection.    

\subsection{Network Architecture and Diffusion Parameters}
We use patch diffusion~\cite{luhman2022improving} for our pre-training, which is an improved version of guided diffusion~\cite{dhariwal2021diffusion}. Patch diffusion proposes to first reshape the image into a grid
of non-overlapping patches and concatenate them in channel dimension, then learn the generation of the reshaped images. This approach reduces the computational requirements for generating high-resolution images. In our experiments, we divide the $512\times512\times1$ images into patches of shape $128\times128\times16$.

In the training of diffusion, we opt for the method of predicting $x_0$, where in each iteration, the network outputs $f_\theta(x_t, t)$ to predict the initial image, as (\ref{eq10__}). 
In terms of network architecture, we employ the U-net architecture from guided diffusion. Some of the network parameters are listed in Table \ref{tab1}, where Dim denotes the number of initial channel dimensions, Muls ($\{m[1],\ldots,m[k]\}$) denotes the channel multipliers, N-Res denotes the number of residual blocks, Res-Attn denotes the resolution at which a self-attention residual block is employed, Dropout denotes the  dropout rate, and Ch-Head denotes the number of channels per attention head.
\begin{table}[thb]\centering
	\caption{Detailed Parameters of the U-Nets Architecture.}
	\label{tab1}
	\resizebox{0.48\textwidth}{!}{
		\Huge
		\begin{tabular}{*{6}{c}}
			\toprule
			 Dim & Muls & \multirow{2}{*}{N-Res} & \multirow{2}{*}{Res-Attn} & \multirow{2}{*}{Dropout} & \multirow{2}{*}{Ch-Head} \\
			 $d$ & $\{m[1],\ldots,m[k]\}$ &  ~ & ~ &  ~ & ~  \\
			\midrule
			 128 & \{1,1,2,2,4\}& 2 & 16, 8 & 0.0 & 64 \\
			\bottomrule
		\end{tabular}
	}
\end{table}

We set the batch size to 16, learning rate to $1.0\times10^{-4}$, and train the model for 150,000 iterations. The training is conducted on a single RTX 3090 GPU.

\subsection{Evaluation Metrics}
For the synthetic DeepLesion dataset, where ground truth is available, we employ peak signal-to-noise ratio (PSNR) and structural similarity index (SSIM) as evaluation metrics. We divide the test images into five groups based on the sizes of the implanted metal masks, from large to small, and calculate average metrics for each group. The grouping is [(2061), (890, 881), (451, 254), (124, 118, 112), (53, 35)]. The metrics are calculated with the HU window [-1000, 4208]. Simultaneously, we showcase the visual effects of different methods in terms of metal artifact reduction for various sizes of metallic implants.  

In the case of real clinical datasets, the absence of ground truth restricts us to using visual assessment as the criterion for evaluating algorithm performance. 
\section{Experiments}
\subsection{Ablation Study}
In this subsection, we validate the effectiveness of different modules proposed in Section \ref{sec2b}, including sinogram inpainting and image fusion. The image fusion module encompasses whether to introduce an initial metal-artifact (MA) image. Furthermore, we explore how to select dynamic weight masks to further enhance algorithm performance.
\subsubsection{Validation of Modules}
We validate the MAR performance of using sinogram inpainting or image fusion strategies separately and using them simultaneously on the synthetic DeepLesion dataset\footnote{Here, we utilized constant $\mathcal{M}_f$ and $\mathcal{M}_y$, where $\mathcal{M}_f = \mathcal{M}(1.0)$, $\mathcal{M}_y = \mathcal{M}(0.8)$.}. We divide image domain fusion into two parts for validation: one represented by (\ref{eq22_}), fusing the prior images from diffusion model (denoted as '+Prior Image' in Table \ref{tab2}; and the other represented by (\ref{eq24_}) and (\ref{eq25_}), involving fusion with the initial metal artifact image (denoted as '+MA Image' in Table \ref{tab2}. Table \ref{tab2} presents the quantitative metrics. 

\begin{table}[thb]\centering
	\caption{Metrics of using different modules in DuDoDp}
	\label{tab2}
	\resizebox{0.48\textwidth}{!}{
		\Huge
		\begin{tabular}{*{6}{c|c|cc|cc}}
			\toprule
			\multirow{2}{*}{Label}&Sino Domain & \multicolumn{2}{c|}{Image Domain} & \multirow{2}{*}{PSNR} & \multirow{2}{*}{SSIM} \\
			\cline{2-4}
			&Sino Inpainting & +Prior Image &  +MA Image & ~ &  ~  \\
			\midrule
			(a)&\Checkmark & \XSolidBrush & \XSolidBrush & 43.04 & 0.988  \\
			\midrule
			(b)&\XSolidBrush & \XSolidBrush & \Checkmark & 40.14 & 0.951  \\
			\midrule
			(c)&\Checkmark & \Checkmark & \XSolidBrush & 43.65 & 0.988  \\
			\midrule
			(d)&\Checkmark & \Checkmark & \Checkmark & \textbf{43.80 }& \textbf{0.989}  \\
			\bottomrule
		\end{tabular}
	}
\end{table}

From Table \ref{tab2}, it can be observed that starting with sinogram inpainting as the foundation achieves a rough reduction of metal artifacts. Image domain fusion further enhances the effectiveness, particularly evident in the improved PSNR. It's evident that each module we proposed has a positive impact.

It's worth noting that we test the effect of solely using the metal artifact image in image domain without sinogram inpainting, as in experiment (b) in Table \ref{tab2}. The process involves iterating according to (\ref{eq24_}) without the fusion of (\ref{eq25_}). The results indicate poorer performance, which can be attributed to the complexity of metal artifacts in the image domain. This demonstrates the necessity of dual-domain processing.

\begin{figure}[htbp]
	\centering\includegraphics[width=0.48\textwidth]{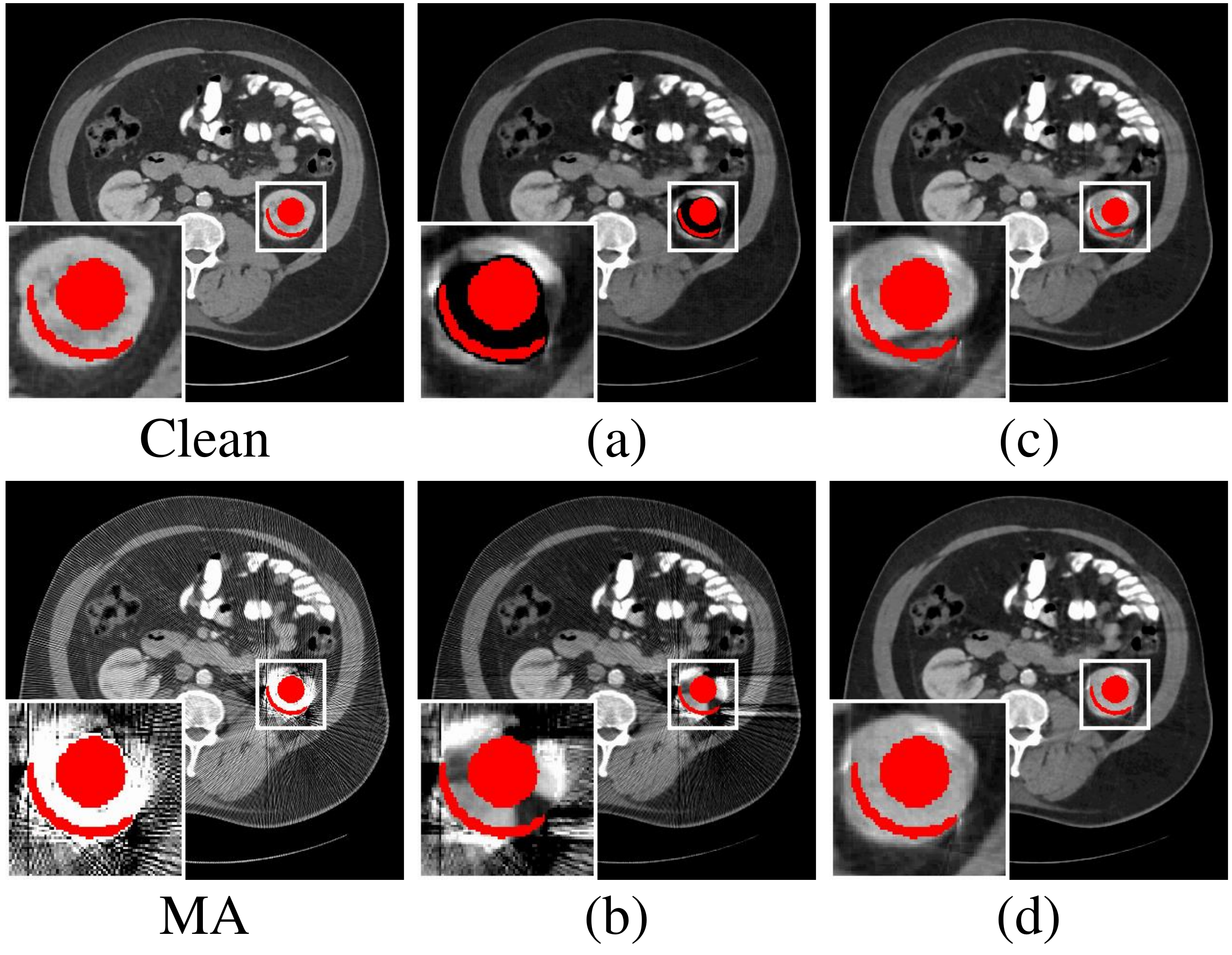}
	\caption{The visual effects of our MAR methods using different modules, where the red parts indicate metallic implants. (a-d) correspond to the experimental setups in Table \ref{tab2}. The display window is [-175, 275] HU.} \label{fig4}
\end{figure}
\begin{figure*}[t]
	\centering\includegraphics[width=0.98\textwidth]{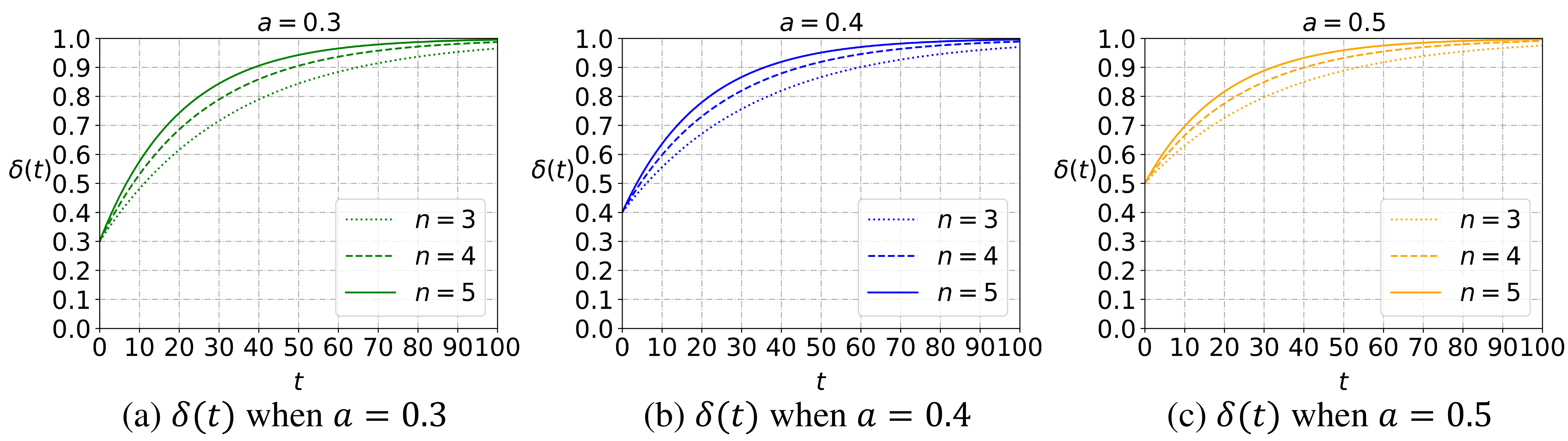}
	\caption{Functions of $\delta(t)$ when $a=0.3, 0.4, 0.5$ and $n=3, 4, 5$.} \label{fig5}
\end{figure*}

From Fig. \ref{fig4}, we can observe that neither using only sinogram domain processing nor image domain processing could successfully recover the image around the metallic implant. However, incorporating image domain fusion after sinogram inpainting leads to improved recovery around the metallic region. The best visual results are achieved when fusing both the diffusion prior image and the original metal artifact image.
\subsubsection{Validation of Dynamic Weight Masks} \label{sec4a2}
In section \ref{secb3}, we introduce the use of temporally dynamic weight masks $\mathcal{M}_f$ to fuse the diffusion prior image and the reconstructed image from inpainted sinogram. Here, we test how to choose the time-varying $\delta(t)$ to generate the dynamic masks, $\mathcal{M}_f(t) = \mathcal{M}(\delta(t))$. Empirically, as the value of timestep $t$ decreases, the artifacts in the image domain become slighter. Hence, we set $\delta(t)$ as a function that positively correlates with timestep $t$,
\begin{equation}
	\delta(t) = (a-1)e^{-n\frac{t}{T}}+1.
\end{equation}
With this design, for a large n value, $\delta(t)$ approaches 1 when $t=T$, while $t=0$, $\delta(t)=a$. Choosing different values of $a$ and $n$ yields different functions, as illustrated in Fig. \ref{fig5}, where we select $a=(0.3, 0.4, 0.5)$ and $n=(3, 4, 5)$ and plot the nine $\delta(t)$ functions obtain from these combinations.

We test the dynamic masks corresponding to the nine $\delta(t)$ functions from Fig. \ref{fig5} on the synthetic DeepLesion dataset. Table \ref{tab3} presents the corresponding PSNR values, where $a=1.0$ mean a constant weight mask because $\delta(t)=1$. 
\begin{table}[thb]\centering
	\caption{PSNRs of using different dynamic weight masks or using constant weight mask}
	\label{tab3}
	\resizebox{0.48\textwidth}{!}{
		\tiny
		\begin{tabular}{*{5}{c|c|ccc}}
			\toprule
			~&~ & $n=3$ &  $n=4$ & $n=5$ \\
			\midrule
			\multirow{3}{*}{Dynamic}&$a=0.3$ & \textbf{43.86} &  43.85 & 43.84  \\
	
			~&$a=0.4$ & \textbf{43.86} &  \textbf{43.86} & 43.85  \\
	
			~&$a=0.5$ & 43.85 &  43.85 & \textbf{43.86}  \\
			\midrule
			Constant&$a=1.0$ & \multicolumn{3}{c}{43.80} \\
			\bottomrule
		\end{tabular}
	}
\end{table}

It's evident that the performance of our method is further enhanced with the use of the proposed dynamic weight masks, compared to the PSNR of 40.80dB with a constant mask. Furthermore, from Table \ref{tab3}, it can be observed that the dynamic mask is quite robust to the parameters $a$ and $n$.
\subsection{Comparison Methods}
We compare our proposed method with various types of MAR methods, including traditional algorithms and deep learning algorithms, unsupervised methods and supervised methods. 

The traditional methods include \textbf{LI}~\cite{kalender1987reduction} and \textbf{NMAR}~\cite{meyer2010normalized} which involve inpainting the metal-affected regions in the sinogram domain.

Unsupervised deep learning-based methods include \textbf{CycleGAN}~\cite{zhu2017unpaired}, \textbf{ADN}~\cite{liao2019adn}, and \textbf{Score-MAR}~\cite{song2022solving}. Among these, CycleGAN and ADN are methods that utilize unpaired clean images and metal artifact images to learn style transfer in the image domain. Although CycleGAN is not specifically designed for MAR, it can be applied to metal artifact reduction. ADN is tailored for MAR and consequently exhibits superior performance. Score-MAR employs a pre-trained score model as a prior for MAR. Its solution primarily involves fusing the score prior with known likelihood in the sinogram domain. This is similar to the sinogram inpainting module in our method, while the main difference lies in its operation not being conducted on the predicted initial image. Furthermore, the original Score-MAR work mainly focuses on introducing a generalized framework for solving inverse problems, hence its experimental conditions for MAR aren't stringent, such as utilizing only parallel beam projections. In the implementation of Score-MAR, to ensure fairness, we employ the same pre-trained model as in our approach\footnote{In theory, score models and diffusion models are equivalent.}. 

Additionally, we also compare two supervised learning methods. The first is \textbf{CNNMAR}~\cite{zhang2018convolutional}, an earlier approach that utilizes a convolutional neural network to predict a prior image for sinogram inpainting. The second is \textbf{InDuDoNet+}~\cite{wang2023indudonet+}, which unfolds the reconstruction process to achieve dual-domain learning. It currently holds the state-of-the-art performance on the synthetic dataset.
\subsection{Comparison on Synthetic Data}
\begin{figure*}[t]
	\centering\includegraphics[width=0.91\textwidth]{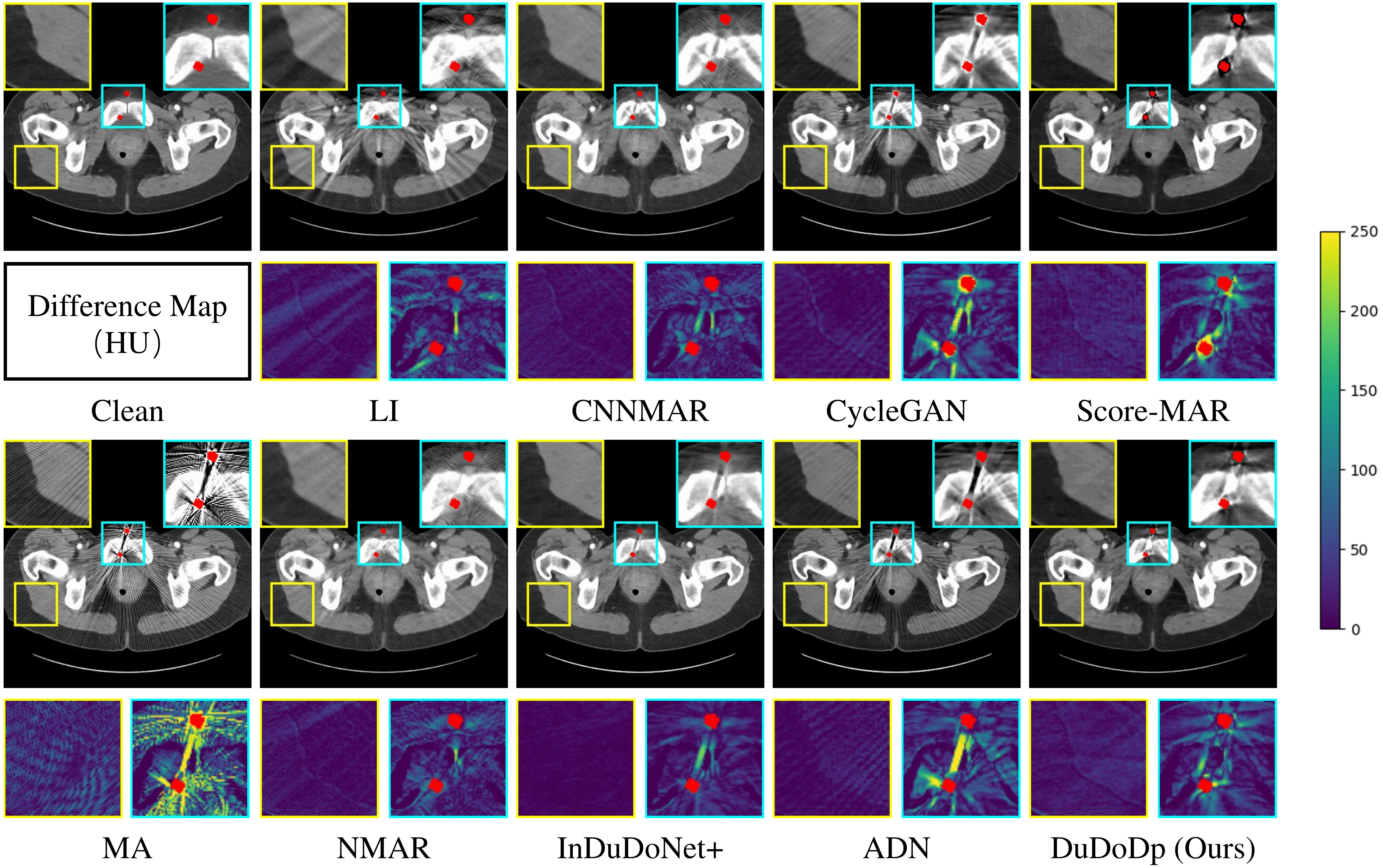}
	\caption{Visual results of different MAR methods on synthetic DeepLesion dataset with a small metallic implant. The red parts indicate metallic implants. The display window is [-175, 275] HU.} \label{fig8_}
\end{figure*}
\begin{table*}[thb]\centering
	\caption{PSNR/SSIM of different MAR methods on synthetic DeepLesion dataset, where bold represents the best result among unsupervised methods, and an asterisk (*) represents the best result including supervised methods.}
	\label{tab4}
	\resizebox{0.98\textwidth}{!}{
		\large
		\begin{tabular}{*{8}{c}}
			\toprule
			~ & Methods&Large metal & \multicolumn{3}{c}{$\rightarrow$} & Small metal & Average \\
			\midrule
			~ & MA & 28.99/0.635 & 30.03/0.766 &  33.05/0.803 & 36.33/0.832 & 37.38/0.853 & 33.89/0.798  \\
			\midrule
			\multirow{2}{*}{Traditional} & LI~\cite{kalender1987reduction} & 37.31/0.894 & 37.62/0.915 &  39.15/0.929 & 41.84/0.943 & 42.36/0.948 & 40.11/0.931  \\
			~ & NMAR~\cite{meyer2010normalized} & 38.16/0.909 & 38.30/0.924 &  39.92/0.934 & 41.80/0.938 & 43.74/0.950 & 40.55/0.934  \\
			\midrule
			\multirow{2}{*}{Supervised} & CNNMAR~\cite{zhang2018convolutional} & 39.29/0.916 & 39.61/0.933 &  41.43/0.945 & 42.92/0.951 & 43.13/0.953 & 41.64/0.943  \\
			~ & InDuDoNet+~\cite{wang2023indudonet+} & $44.17^*$/$0.991^*$ & $45.52^*$/$0.994^*$ &  $48.63^*$/$0.995^*$ & $53.87^*$/$0.998^*$ & $54.63^*$/$0.998^*$ & $50.33^*$/$0.996^*$  \\
			\midrule
			\multirow{4}{*}{Unsupervised} & CycleGAN~\cite{zhu2017unpaired} & 36.04/0.957 & 38.62/0.974 &  41.00/0.977 & 43.44/0.983 & 44.19/0.985 & 41.40/0.978  \\
			~ & ADN~\cite{liao2019adn} & 36.43/0.966 & 41.76/0.984 &  42.22/0.985 & \textbf{45.38}/0.989 & \textbf{46.13}/0.990 & 43.28/0.985  \\
			~ & Score-MAR~\cite{song2022solving} & 41.73/0.985 & 41.35/0.985 &  41.35/0.985 & 42.58/0.987 & 44.21/0.990 & 43.04/0.988  \\
			~ & DuDoDp (Ours) & \textbf{42.81}/\textbf{0.986} & \textbf{42.43}/\textbf{0.986} &  \textbf{43.42}/\textbf{0.988} & 44.78/\textbf{0.990} & 44.89/\textbf{0.991} & \textbf{43.86}/\textbf{0.989}  \\
			\bottomrule
		\end{tabular}
	}
\end{table*}
\begin{figure*}[t]
	\centering\includegraphics[width=0.91\textwidth]{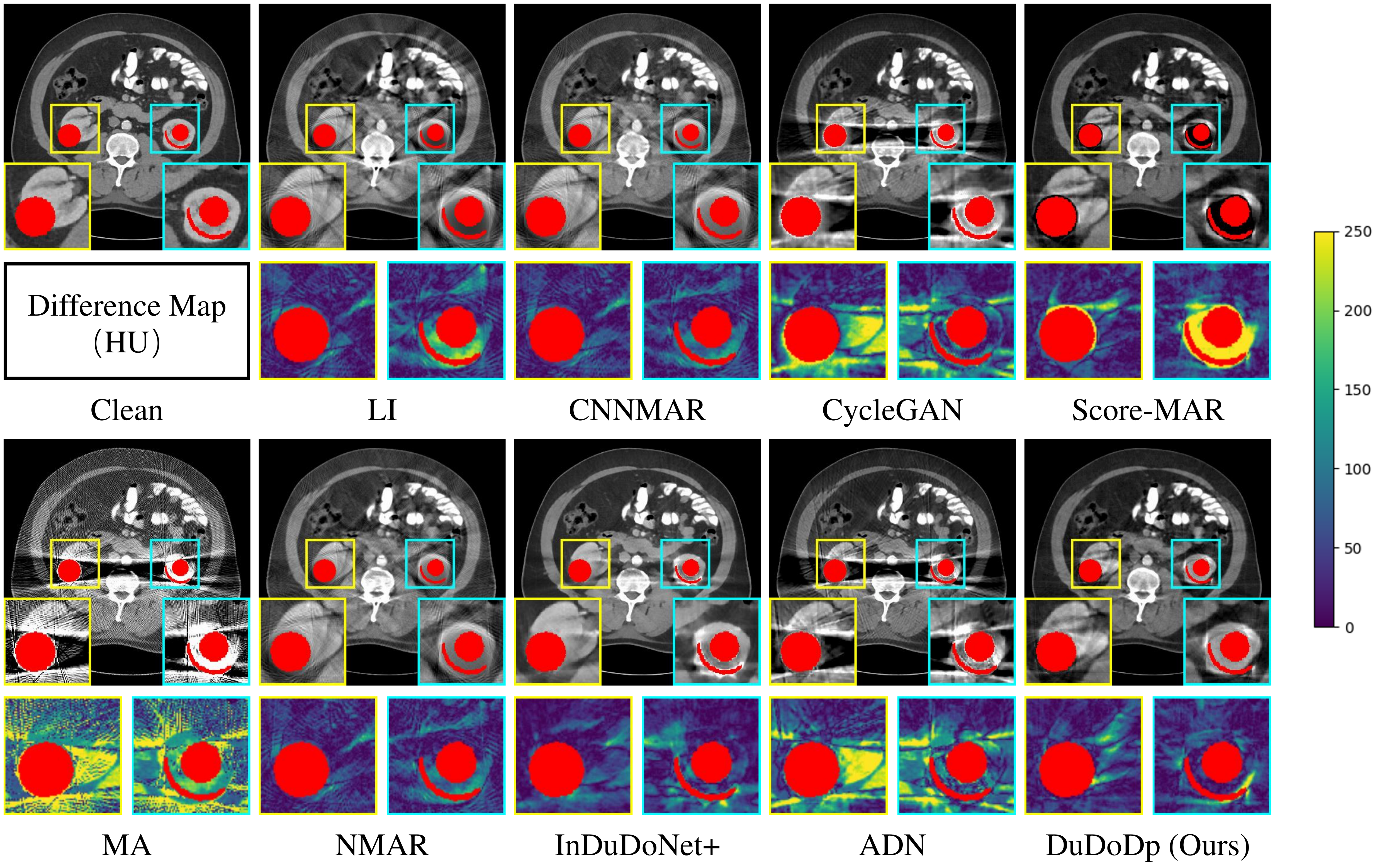}
	\caption{Visual results of different MAR methods on synthetic DeepLesion dataset with large metallic implants. The red parts indicate metallic implants. The display window is [-175, 275] HU.} \label{fig9_}
\end{figure*}
\begin{figure*}[t]
	\centering\includegraphics[width=0.91\textwidth]{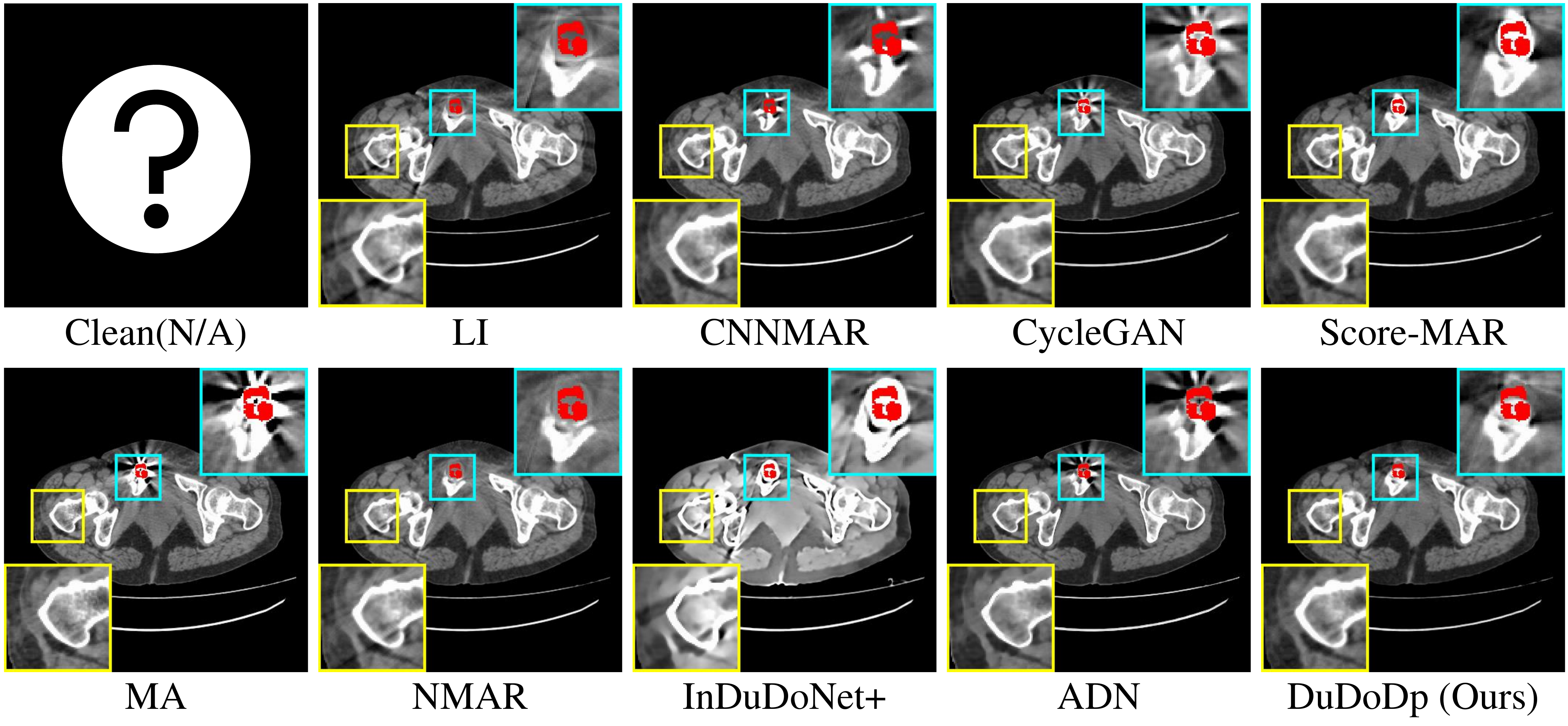}
	\caption{Visual results of different MAR methods on one CT image from clinical CTPelvic1K dataset. The red parts indicate metallic implants. The display window is [-175, 275] HU.} \label{fig7}
\end{figure*}

We first conduct the comparison methods and our method on the synthetic DeepLesion dataset. The metrics of the MAR results are shown in Table \ref{tab4}. Based on the size of the metallic implants, the test data are divided into five groups to calculate average metrics for each group. Finally, the overall average metrics are also calculated. From Table \ref{tab4}, it can be observed that our method achieves the best overall average performance among unsupervised algorithms, and surpassing the earlier supervised method CNNMAR, but still with a certain gap compared to the state-of-the-art supervised method InDuDoNet+. Among the unsupervised methods, CycleGAN and ADN, which operate solely in the image domain, perform poorly on large metallic implants but perform better on smaller ones. Our method, benefiting from dual-domain information, performs better on large metal artifacts. Score-MAR's limited performance is attributed to its use of score priors only for sinogram inpainting. This highlights the advantages of using the diffusion priors in dual domains.  

Fig. \ref{fig8_} and Fig. \ref{fig9_} display the visual effects of different methods' MAR results. As Fig. \ref{fig8_} and Fig. \ref{fig9_} show, metallic implants can introduce severe streak artifacts in CT images. While LI and NMAR significantly mitigate metal artifacts by performing inpainting in the sinogram domain, they introduce new artifacts due to the discontinuity between inpainted and known parts of the sinogram. CNNMAR improves upon NMAR using convolutional neural networks, but its results still exhibit some artifacts like NMAR. InDuDoNet+ achieves excellent results by supervised learning in dual domains, yet residual streak artifacts are visible in its output image, especially as shown in Fig. \ref{fig9_}.

CycleGAN and ADN employ unsupervised learning in the image domain to map metal artifact images to clean images. However, due to the complexity and severity of metal artifacts, as seen in Fig. \ref{fig8_} and Fig. \ref{fig9_}, they struggle to effectively remove streak artifacts, especially with larger metallic implants.

Both Score-MAR and our method utilize deep generative models as priors. Their results appear more visually realistic than other methods. However, Score-MAR solely uses the score priors for sinogram inpainting, which leads to substantial errors around metallic implants. In contrast, our method, through iterative dual-domain processing, restores a natural and realistic appearance around the metallic implants.
\begin{table*}[h]\centering 
	\captionof{table}{Radiologist evaluation results on clinical data.}
	\label{tab5_}
	\resizebox{0.7\textwidth}{!}{
		\tiny
		\begin{tabular}{c|ccc|ccc}
			\toprule
			\multirow{2}{*}{Method} & \multicolumn{3}{c|}{Radiologist 1} & \multicolumn{3}{c}{Radiologist 2} \\
			~ &  Top-1  & Top-2 & Worst-1 & Top-1  & Top-2 & Worst-1\\
			\midrule
			LI~\cite{kalender1987reduction}  & 0\%  & 0\% & \textbf{48\%} & 0\% & 6\% & \textbf{46\%}\\
			NMAR~\cite{meyer2010normalized}  & 4\%  & 20\% & 2\% & 8\% & 28\% & 4\%\\
			CNNMAR~\cite{zhang2018convolutional} & 14\%  & 48\% & 0\% & 16\% & 54\% & 0\%\\
			InDuDoNet+~\cite{wang2023indudonet+}  & 0\%  & 0\% & 18\% & 0\% & 0\% & 26\%\\
			CycleGAN~\cite{zhu2017unpaired}  & 2\%  & 4\% & 8\% & 0\% & 2\% & 6\%\\
			ADN~\cite{liao2019adn}  & 2\%  & 6\% & 18\% & 0\% & 4\% & 18\%\\
			Score-MAR~\cite{song2022solving}  & 16\%  & 40\% & 6\% & 18\% & 30\% & 0\%\\
			DuDoDp (Ours) & \textbf{62\%}  & \textbf{82\%} & 0\% & \textbf{58\%} & \textbf{76\%} & 0\%\\
			
			\bottomrule
		\end{tabular} 
	}
\end{table*}

From the difference maps, it can be observed that our method outperforms other unsupervised deep learning methods, both for large and small metallic implants. Surprisingly, although the overall visual effects are limited, traditional methods LI and NMAR exhibit excellent difference maps in the neighbor of metallic implants. However, they introduce new artifacts in areas away from the metal, as indicated by the yellow ROI (region of interest) in Fig. \ref{fig8_}. These introduced artifacts adversely affect both the visual quality and metrics of the entire image. In contrast, our method successfully avoids this issue. The difference maps of supervised methods, overall, show good performance. Our method achieves results close to the state-of-the-art supervised methods even without using paired training data.

\subsection{Comparison on clinical Data}
Fig. \ref{fig7} shows the results of different methods on the clinical dataset. The images depict a pelvic CT slice. Due to the presence of metallic implants, significant artifacts are observed around the metal regions. However, the impact on areas farther from the metallic implants is relatively minor. Visually, traditional methods LI and NMAR have achieved promising outcomes, as indicated by the light blue region of interest, where they manage to achieve good restoration around the metal areas. However, LI and NMAR introduce new artifacts in regions away from the metallic implants, as demonstrated by the yellow ROI. CNNMAR, with the assistance of convolutional neural networks, significantly mitigates the shortcomings of traditional methods. However, near metal implants, CNNMAR introduces some spurious artifacts.

Purely image-based unsupervised methods, including CycleGAN and ADN, do not perform well on clinical data, as significant artifacts persist around metal regions. This could arise from the learned image domain translation not aligning well with the clinical data due to domain gaps. Moreover, these methods are unable to fully leverage the information present in the sinogram. While they don't introduce new artifacts in the yellow ROI region, they also fail to remove the artifacts near the metal implants.

The results of InDuDoNet+ on clinical data are intriguing. It successfully eliminates artifacts around metal areas, but overall, its results slightly alter the overall brightness of CT images. In particular, the contrast of the bone becomes higher. This may explain the excellent performance of pelvic fracture segmentation tasks using InDuDoNet's MAR results~\cite{wang2021indudonet}. However, compared to earlier supervised learning algorithm CNNMAR, InDuDoNet+, while demonstrating a significant performance advantage on simulated datasets, exhibits weaker visual effects on clinical data. This suggests that an overly complex network may lead to overfitting on training datasets, potentially compromising the generalization of the method.

Score-MAR partially mitigates the metal artifacts, yet there remains an unnatural region around the metallic implant. In comparison to the aforementioned methods, our approach not only successfully reduces the artifacts around the metal but also refrains from introducing new artifacts in other regions. Considering these aspects, our method exhibits the most favorable visual results among all comparison methods, highlighting its potential applicability in clinical settings.

Moreover, We conducted experiments on 50 metal artifact CT images from the CTPelvic1K dataset. The results were randomly shuffled, and the method names were concealed to create a questionnaire. We invited two experienced radiology experts to select the best two images and the worst one from each set of results. The statistic results are shown in Table \ref{tab5_}.
	
As shown in the table, our method's results are more favored by professional radiologists. It ranks first in both the best and second-best selection rates and rarely produces very poor results. Regarding other methods, it is interesting to note that among supervised algorithms, CNNMAR, which utilizes a small model to combine with traditional method, performs significantly better on clinical data than the state-of-the-art InDuDoNet+. This suggests that complex network structures and excessive training can harm the generalization of supervised methods, despite their outstanding performance on synthetic datasets.

\begin{table*}\centering 
	\caption{Inference time of our method with/without acceleration. The tests are conducted on an Nvidia RTX 3090 GPU.}
	\label{tab6_}
	\resizebox{0.9\textwidth}{!}{
		\large
		\begin{tabular}{*{6}{c}}
			\toprule
			& Number of Timesteps & Batch Size & Memory(G) & Sum of Time(s) & Time per Image(s) \\
			\midrule
			\multirow{3}{*}{No Acceleration}& 1000 & 1 & 1.82& 19.18  & 19.18 \\
			~& 1000 & 4 &2.04&  36.85 & 9.21 \\
			~& 1000 & 16 &3.50& 129.36  & 8.09 \\
			~& 100 & 64 & 9.16& 486.85  & 7.61 \\
			\midrule
			\multirow{4}{*}{Acceleration} & 100 & 1 &1.82&  1.83 & 1.83 \\
			~& 100 & 4 & 2.04 & 3.46  & 0.87 \\
			~& 100 & 16 & 3.50& 12.19  & 0.76 \\
			~& 100 & 64 & 9.16&  47.82 & 0.75 \\
			\bottomrule
		\end{tabular} 
	}
\end{table*}

\begin{table*}\centering 
	\caption{Model size and computational efficiency of deep learning-based MAR methods. The tests are conducted on an Nvidia RTX 3090 GPU. All of the networks are test with batch size 1.}
	\label{tab7_}
	\resizebox{\textwidth}{!}{
		\small
		\begin{tabular}{*{7}{c}}
			\toprule
			Method & CycleGAN~\cite{zhu2017unpaired} & ADN~\cite{liao2019adn}  & Score-MAR~\cite{song2022solving} & CNNMAR~\cite{zhang2018convolutional} & InDuDoNet+~\cite{wang2023indudonet+} & DuDoDp (Ours) \\
			\midrule
			Parameters (M) &  11.37  & 22.43 & 80.11 & 0.03 & 1.75 & 80.11\\
			FLOPs (G) &  122.43 & 375.32  & 130.54 & 5.03 & 510.74 & 130.54\\
			Inference time (s)  & 0.09  & 0.12  & 1.77 & 0.01 & 0.37 &  1.83\\
			\bottomrule
		\end{tabular} 
	}
\end{table*}

\subsection{Comparison of model size and inference time}
Due to the iterative nature of the diffusion model, inference time is often a concern. In our method, to reduce the time required for the algorithm, we use the method from~\cite{song2020denoising} to reduce the number of inference steps from 1000 to 100. The comparison of inference time with or without acceleration is shown in Table \ref{tab6_}. 

As we can see, after using acceleration technique, the inference time is acceptable. Besides, by using parallel computing (calculating a batch of images at the same time), the average inference time per image can be reduced.
	
In addition, we tested the model parameters, FLOPs (Floating Point Operations), and the inference time required for various deep learning-based MAR methods. The results are presented in Table \ref{tab7_}.
	
From Table \ref{tab7_}, it can be seen that our method utilizes the largest model. This is because, as an unconditional generative model, a larger model is necessary to ensure the capture of prior information across the entire CT image domain. At the same time, our inference time is also the longest since our approach operates iteratively. The FLOPs of our method are not the highest, thanks to the high inference efficiency of the U-Nets.

\section{Discussion and Conclusion}

\subsection{Limitation}
One of the major limitations of our method is the computational time. For each metal artifact reduction process, our approach requires 100 iterations, with each iteration involving an inference of the pre-trained U-Net model. This results in significantly longer computation times compared to other end-to-end deep learning-based MAR methods, potentially limiting the clinical applicability of our approach. Therefore, future research on accelerating the diffusion model may benefit the efficacy of our MAR method.
	
Another noteworthy issue is the performance in handling small metal artifacts. Our method does not outperform the ADN method, which is trained on unpaired clean CT images and metal artifact CT images, even though ADN is a purely image-domain method. This suggests that our method may lack likelihood modeling in the image domain, as we chose to model the MAR problem as an inpainting problem in the projection domain. It's worth exploring whether improving the utilization of image-domain information can be achieved by introducing additional training conditions like ADN or incorporating image-domain metal artifact models. This direction is worth discussing to enhance the performance of our method.

\subsection{Discussion}
A potential way for further enhancing our algorithm's performance lies in adopting distinct image fusion strategies for varying sizes of metallic implants. As evident from Table \ref{tab4}, in experiments involving small metallic implants, our algorithm's metrics are inferior to those of ADN, a pure image-domain algorithm. This observation suggests that when dealing with smaller metallic implants, the metal artifact images still contain substantial clean image information. Consequently, an increased addition of metal artifact images during the image fusion stage could potentially lead to a further improvement in algorithm performance.

We introduce dynamic weight masks to incorporate the diffusion priors in the image domain. The dynamic masks are designed to rectify the newly introduced artifacts after inpainted sinogram reconstruction and they correlate with the timestep $t$. Despite the enhancement offered by the dynamic masks, its scope remains somewhat limited. Therefore, We consider that these weight masks can also be content-dependent, resulting in different masks for different images. Such an approach would enhance the adaptability of our algorithm and is an aspect worth investigating.  
 
Prompted by the above two points, the question arises: how can we derive the most suitable weight masks based on image content, shape of the metallic implant, and timestep  $t$? Manual design of a function is inherently suboptimal in this context. Therefore, we propose the idea of utilizing paired data to learn the image fusion, transforming our method into a model-based supervised learning approach. Although this deviates somewhat from our original motivation, it's foreseeable that such an approach could significantly enhance the performance of our algorithm. This might position our method to be comparable with state-of-the-art supervised learning algorithms like InDuDoNet+. It remains an avenue to be explored in the future.
\subsection{Conclusion}
In this paper, we propose DuDoDp for metal artifact reduction, a dual domain method with diffusion priors. Firstly, we train a denoising diffusion probabilistic model with clean CT images. To introduce the diffusion priors to the solving of MAR, we formulate an optimization problem at each timestep of the diffusion model, which solution corresponds to a sinogram inpainting module. After, we propose to fuse the reconstructed image of the inpainted sinogram, the diffusion prior image, and the metal artifact image to make better use of the image domain information. To better control the weights of image fusion, we design dynamic weight masks to further enhance performance. The ablation studies show that the modules proposed in our approach positively contribute to the results of MAR. The results on both the synthetic and clinical datasets have demonstrated that our method outperforms all other unsupervised methods and surpasses earlier supervised MAR method. Our method shows promise in clinical applications, especially in the presence of larger metallic implants such as hip prostheses. However, for smaller metal objects, certain image-domain methods trained on unpaired images, such as ADN, may yield better results. Therefore, the application of our method in tasks of artifact reduction for small metallic objects, such as metal coils, needs to be carefully considered. We believe that our work holds significant implications in exploring the application of deep generative models in the field of MAR. This could potentially extend the use of metal artifact reduction techniques to scenarios where paired training data is scarce.  
\section*{Acknowledgment}
We express our gratitude to the codes of PatchDiffusion-Pytorch~\cite{luhman2022improving} and guided-diffusion~\cite{dhariwal2021diffusion} that facilitate the pre-training of our diffusion model. 

\bibliographystyle{IEEEtran}
\bibliography{IEEEabrv,tmi2023}

\begin{thebibliography}{10}
\providecommand{\url}[1]{#1}
\csname url@samestyle\endcsname
\providecommand{\newblock}{\relax}
\providecommand{\bibinfo}[2]{#2}
\providecommand{\BIBentrySTDinterwordspacing}{\spaceskip=0pt\relax}
\providecommand{\BIBentryALTinterwordstretchfactor}{4}
\providecommand{\BIBentryALTinterwordspacing}{\spaceskip=\fontdimen2\font plus
\BIBentryALTinterwordstretchfactor\fontdimen3\font minus
  \fontdimen4\font\relax}
\providecommand{\BIBforeignlanguage}[2]{{%
\expandafter\ifx\csname l@#1\endcsname\relax
\typeout{** WARNING: IEEEtran.bst: No hyphenation pattern has been}%
\typeout{** loaded for the language `#1'. Using the pattern for}%
\typeout{** the default language instead.}%
\else
\language=\csname l@#1\endcsname
\fi
#2}}
\providecommand{\BIBdecl}{\relax}
\BIBdecl

\bibitem{de1998metal}
B.~De~Man, J.~Nuyts, P.~Dupont, G.~Marchal, and P.~Suetens, ``Metal streak
  artifacts in x-ray computed tomography: a simulation study,'' in \emph{1998
  IEEE Nuclear Science Symposium Conference Record. 1998 IEEE Nuclear Science
  Symposium and Medical Imaging Conference (Cat. No. 98CH36255)}, vol.~3.\hskip
  1em plus 0.5em minus 0.4em\relax IEEE, 1998, pp. 1860--1865.

\bibitem{park2018ct}
H.~S. Park, S.~M. Lee, H.~P. Kim, J.~K. Seo, and Y.~E. Chung, ``Ct
  sinogram-consistency learning for metal-induced beam hardening correction,''
  \emph{Medical physics}, vol.~45, no.~12, pp. 5376--5384, 2018.

\bibitem{kalender1987reduction}
W.~A. Kalender, R.~Hebel, and J.~Ebersberger, ``Reduction of ct artifacts
  caused by metallic implants.'' \emph{Radiology}, vol. 164, no.~2, pp.
  576--577, 1987.

\bibitem{meyer2010normalized}
E.~Meyer, R.~Raupach, M.~Lell, B.~Schmidt, and M.~Kachelrie{\ss}, ``Normalized
  metal artifact reduction (nmar) in computed tomography,'' \emph{Medical
  physics}, vol.~37, no.~10, pp. 5482--5493, 2010.

\bibitem{lin2019dudonet}
W.-A. Lin, H.~Liao, C.~Peng, X.~Sun, J.~Zhang, J.~Luo, R.~Chellappa, and S.~K.
  Zhou, ``Dudonet: Dual domain network for ct metal artifact reduction,'' in
  \emph{Proceedings of the IEEE/CVF Conference on Computer Vision and Pattern
  Recognition}, 2019, pp. 10\,512--10\,521.

\bibitem{wang2021indudonet}
H.~Wang, Y.~Li, H.~Zhang, J.~Chen, K.~Ma, D.~Meng, and Y.~Zheng, ``Indudonet:
  An interpretable dual domain network for ct metal artifact reduction,'' in
  \emph{Medical Image Computing and Computer Assisted Intervention--MICCAI
  2021: 24th International Conference, Strasbourg, France, September
  27--October 1, 2021, Proceedings, Part VI 24}.\hskip 1em plus 0.5em minus
  0.4em\relax Springer, 2021, pp. 107--118.

\bibitem{wang2022ada}
H.~Wang, Y.~Li, D.~Meng, and Y.~Zheng, ``Adaptive convolutional dictionary
  network for ct metal artifact reduction,'' in \emph{The 31st International
  Joint Conference on Artificial Intelligence}.\hskip 1em plus 0.5em minus
  0.4em\relax IEEE, 2022.

\bibitem{zhou2022dudodr}
B.~Zhou, X.~Chen, S.~K. Zhou, J.~S. Duncan, and C.~Liu, ``Dudodr-net:
  Dual-domain data consistent recurrent network for simultaneous sparse view
  and metal artifact reduction in computed tomography,'' \emph{Medical Image
  Analysis}, vol.~75, p. 102289, 2022.

\bibitem{wang2023indudonet+}
H.~Wang, Y.~Li, H.~Zhang, D.~Meng, and Y.~Zheng, ``Indudonet+: A deep unfolding
  dual domain network for metal artifact reduction in ct images,''
  \emph{Medical Image Analysis}, vol.~85, p. 102729, 2023.

\bibitem{zhang2018convolutional}
Y.~Zhang and H.~Yu, ``Convolutional neural network based metal artifact
  reduction in x-ray computed tomography,'' \emph{IEEE transactions on medical
  imaging}, vol.~37, no.~6, pp. 1370--1381, 2018.

\bibitem{huang2018metal}
X.~Huang, J.~Wang, F.~Tang, T.~Zhong, and Y.~Zhang, ``Metal artifact reduction
  on cervical ct images by deep residual learning,'' \emph{Biomedical
  engineering online}, vol.~17, pp. 1--15, 2018.

\bibitem{goodfellow2020generative}
I.~Goodfellow, J.~Pouget-Abadie, M.~Mirza, B.~Xu, D.~Warde-Farley, S.~Ozair,
  A.~Courville, and Y.~Bengio, ``Generative adversarial networks,''
  \emph{Communications of the ACM}, vol.~63, no.~11, pp. 139--144, 2020.

\bibitem{wang2018conditional}
J.~Wang, Y.~Zhao, J.~H. Noble, and B.~M. Dawant, ``Conditional generative
  adversarial networks for metal artifact reduction in ct images of the ear,''
  in \emph{Medical Image Computing and Computer Assisted Intervention--MICCAI
  2018: 21st International Conference, Granada, Spain, September 16-20, 2018,
  Proceedings, Part I}.\hskip 1em plus 0.5em minus 0.4em\relax Springer, 2018,
  pp. 3--11.

\bibitem{liao2019adn}
H.~Liao, W.-A. Lin, S.~K. Zhou, and J.~Luo, ``Adn: artifact disentanglement
  network for unsupervised metal artifact reduction,'' \emph{IEEE Transactions
  on Medical Imaging}, vol.~39, no.~3, pp. 634--643, 2019.

\bibitem{lyu2021u}
Y.~Lyu, J.~Fu, C.~Peng, and S.~K. Zhou, ``U-dudonet: unpaired dual-domain
  network for ct metal artifact reduction,'' in \emph{Medical Image Computing
  and Computer Assisted Intervention--MICCAI 2021: 24th International
  Conference, Strasbourg, France, September 27--October 1, 2021, Proceedings,
  Part VI 24}.\hskip 1em plus 0.5em minus 0.4em\relax Springer, 2021, pp.
  296--306.

\bibitem{du2023deep}
M.~Du, K.~Liang, L.~Zhang, H.~Gao, Y.~Liu, and Y.~Xing, ``Deep-learning-based
  metal artefact reduction with unsupervised domain adaptation regularization
  for practical ct images,'' \emph{IEEE Transactions on Medical Imaging}, 2023.

\bibitem{zhu2017unpaired}
J.-Y. Zhu, T.~Park, P.~Isola, and A.~A. Efros, ``Unpaired image-to-image
  translation using cycle-consistent adversarial networks,'' in
  \emph{Proceedings of the IEEE international conference on computer vision},
  2017, pp. 2223--2232.

\bibitem{huang2018multimodal}
X.~Huang, M.-Y. Liu, S.~Belongie, and J.~Kautz, ``Multimodal unsupervised
  image-to-image translation,'' in \emph{Proceedings of the European conference
  on computer vision (ECCV)}, 2018, pp. 172--189.

\bibitem{wu2023unsupervised}
Q.~Wu, L.~Chen, C.~Wang, H.~Wei, S.~K. Zhou, J.~Yu, and Y.~Zhang,
  ``Unsupervised polychromatic neural representation for ct metal artifact
  reduction,'' 2023.

\bibitem{song2021scorebased}
\BIBentryALTinterwordspacing
Y.~Song, J.~Sohl-Dickstein, D.~P. Kingma, A.~Kumar, S.~Ermon, and B.~Poole,
  ``Score-based generative modeling through stochastic differential
  equations,'' in \emph{International Conference on Learning Representations},
  2021. [Online]. Available: \url{https://openreview.net/forum?id=PxTIG12RRHS}
\BIBentrySTDinterwordspacing

\bibitem{song2022solving}
\BIBentryALTinterwordspacing
Y.~Song, L.~Shen, L.~Xing, and S.~Ermon, ``Solving inverse problems in medical
  imaging with score-based generative models,'' in \emph{International
  Conference on Learning Representations}, 2022. [Online]. Available:
  \url{https://openreview.net/forum?id=vaRCHVj0uGI}
\BIBentrySTDinterwordspacing

\bibitem{ho2020denoising}
J.~Ho, A.~Jain, and P.~Abbeel, ``Denoising diffusion probabilistic models,''
  \emph{Advances in Neural Information Processing Systems}, vol.~33, pp.
  6840--6851, 2020.

\bibitem{kingma2013auto}
D.~P. Kingma and M.~Welling, ``Auto-encoding variational bayes,'' \emph{arXiv
  preprint arXiv:1312.6114}, 2013.

\bibitem{dinh2014nice}
L.~Dinh, D.~Krueger, and Y.~Bengio, ``Nice: Non-linear independent components
  estimation,'' \emph{arXiv preprint arXiv:1410.8516}, 2014.

\bibitem{dinh2016density}
L.~Dinh, J.~Sohl-Dickstein, and S.~Bengio, ``Density estimation using real
  nvp,'' \emph{arXiv preprint arXiv:1605.08803}, 2016.

\bibitem{2018Glow}
D.~P. Kingma and P.~Dhariwal, ``Glow: Generative flow with invertible 1x1
  convolutions,'' \emph{Advances in neural information processing systems},
  vol.~31, 2018.

\bibitem{croitoru2023diffusion}
F.-A. Croitoru, V.~Hondru, R.~T. Ionescu, and M.~Shah, ``Diffusion models in
  vision: A survey,'' \emph{IEEE Transactions on Pattern Analysis and Machine
  Intelligence}, 2023.

\bibitem{song2020denoising}
J.~Song, C.~Meng, and S.~Ermon, ``Denoising diffusion implicit models,''
  \emph{arXiv preprint arXiv:2010.02502}, 2020.

\bibitem{hsieh2000iterative}
J.~Hsieh, R.~C. Molthen, C.~A. Dawson, and R.~H. Johnson, ``An iterative
  approach to the beam hardening correction in cone beam ct,'' \emph{Medical
  physics}, vol.~27, no.~1, pp. 23--29, 2000.

\bibitem{kachelriess2001generalized}
M.~Kachelriess, O.~Watzke, and W.~A. Kalender, ``Generalized multi-dimensional
  adaptive filtering for conventional and spiral single-slice, multi-slice, and
  cone-beam ct,'' \emph{Medical physics}, vol.~28, no.~4, pp. 475--490, 2001.

\bibitem{zhang2010beam}
Y.~Zhang, X.~Mou, and S.~Tang, ``Beam hardening correction for fan-beam ct
  imaging with multiple materials,'' in \emph{IEEE Nuclear Science Symposuim \&
  Medical Imaging Conference}.\hskip 1em plus 0.5em minus 0.4em\relax IEEE,
  2010, pp. 3566--3570.

\bibitem{park2015metal}
H.~S. Park, D.~Hwang, and J.~K. Seo, ``Metal artifact reduction for
  polychromatic x-ray ct based on a beam-hardening corrector,'' \emph{IEEE
  transactions on medical imaging}, vol.~35, no.~2, pp. 480--487, 2015.

\bibitem{yan2018deep}
K.~Yan, X.~Wang, L.~Lu, L.~Zhang, A.~P. Harrison, M.~Bagheri, and R.~M.
  Summers, ``Deep lesion graphs in the wild: relationship learning and
  organization of significant radiology image findings in a diverse large-scale
  lesion database,'' in \emph{Proceedings of the IEEE Conference on Computer
  Vision and Pattern Recognition}, 2018, pp. 9261--9270.

\bibitem{yu2020deep}
L.~Yu, Z.~Zhang, X.~Li, and L.~Xing, ``Deep sinogram completion with image
  prior for metal artifact reduction in ct images,'' \emph{IEEE transactions on
  medical imaging}, vol.~40, no.~1, pp. 228--238, 2020.

\bibitem{liu2021deep}
P.~Liu, H.~Han, Y.~Du, H.~Zhu, Y.~Li, F.~Gu, H.~Xiao, J.~Li, C.~Zhao, L.~Xiao
  \emph{et~al.}, ``Deep learning to segment pelvic bones: large-scale ct
  datasets and baseline models,'' \emph{International Journal of Computer
  Assisted Radiology and Surgery}, vol.~16, pp. 749--756, 2021.

\bibitem{luhman2022improving}
T.~Luhman and E.~Luhman, ``Improving diffusion model efficiency through
  patching,'' \emph{arXiv preprint arXiv:2207.04316}, 2022.

\bibitem{dhariwal2021diffusion}
P.~Dhariwal and A.~Nichol, ``Diffusion models beat gans on image synthesis,''
  \emph{Advances in neural information processing systems}, vol.~34, pp.
  8780--8794, 2021.

\end{thebibliography}

\end{document}